\documentclass[aps,amsmath,tightlines,twocolumn,pra,superscriptaddress,nofootinbib]{revtex4}

\usepackage{amssymb}
\usepackage[dvips]{graphicx}
\usepackage{amsmath,psfrag}
\usepackage{psfrag}
\usepackage{longtable}
\usepackage{dcolumn,epsfig}
\usepackage[usenames]{color}
\usepackage[normalem]{ulem}

\begin{document}
\newcommand{\comment}[1]{}
\newcommand{\E}{\mathrm{E}}
\newcommand{\Var}{\mathrm{Var}}
\newcommand{\bra}[1]{\langle #1|}
\newcommand{\ket}[1]{|#1\rangle}
\newcommand{\braket}[2]{\langle #1|#2 \rangle}
\newcommand{\be}{\begin{equation}}
\newcommand{\ee}{\end{equation}}
\newcommand{\ba}{\begin{eqnarray}}
\newcommand{\ea}{\end{eqnarray}}
\newcommand{\SD}[1]{{\color{magenta}#1}}
\newcommand{\HME}[1]{{\color{green}#1}}
\newcommand{\rem}[1]{{\color{cyan} \sout{#1}}}
\newcommand{\alert}[1]{\textbf{\color{red} \uwave{#1}}}
\newcommand{\Y}[1]{\textcolor{BurntOrange}{#1}}
\newcommand{\R}[1]{\textcolor{red}{#1}}
\newcommand{\B}[1]{\textcolor{blue}{#1}}
\title{Probing macroscopic quantum states with a sub-Heisenberg accuracy}

\author{Haixing Miao}
\affiliation{School of Physics, University of Western Australia,
WA 6009, Australia}
\author{Stefan Danilishin}
\affiliation{Physics Faculty, Moscow State University, Moscow
119991, Russia} \affiliation{Max-Planck Institut f\"ur
Gravitationsphysik (Albert-Einstein-Institut) and Leibniz
Universit\"at Hannover, Callinstr. 38, 30167 Hannover, Germany}
\author{Helge M\"uller-Ebhardt}
\author{Henning Rehbein}
\affiliation{Max-Planck Institut f\"ur Gravitationsphysik
(Albert-Einstein-Institut) and Leibniz Universit\"at Hannover,
Callinstr. 38, 30167 Hannover, Germany}
\author{Kentaro Somiya}
\author{Yanbei Chen}
\affiliation{Theoretical Astrophysics 130-33, California Institute
of Technology, Pasadena, CA 91125, USA}

\begin{abstract}
Significant achievements in high-sensitivity measurements will soon allow us
to probe quantum behaviors of macroscopic mechanical oscillators. In a recent
work [arXiv: 0903.0079], we formulated a general framework for treating preparation of
Gaussian quantum states of macroscopic oscillators through linear position measurements. To
outline a complete procedure for testing macroscopic quantum mechanics, here
we consider a subsequent verification stage which probes the prepared macroscopic
quantum state and verifies the quantum dynamics. By adopting an optimal time-dependent
homodyne detection in which the phase of the local oscillator varies in time, the
conditional quantum state can be characterized below the Heisenberg limit, thereby
achieving a quantum tomography. In the limiting case of no readout loss,
such a scheme evades measurement-induced back action, which is identical to the
variational-type measurement scheme invented by Vyatchanin {\it et al.} but in the
context for detecting gravitational waves. To motivate macroscopic quantum mechanics
experiments with future gravitational-wave detectors, we mostly focus on the parameter
regime where the characteristic measurement frequency is much higher than the oscillator
frequency and the classical noises
are Markovian, which captures the main feature of a broadband gravitational-wave
detector. In addition, we discuss verifications of Einstein-Podolsky-Rosen-type
entanglement between macroscopic test masses in future gravitational-wave detectors,
which enables us to test one particular version of gravity decoherence
conjectured by Di\'{o}si and Penrose.

\end{abstract}
\maketitle

\section{Introduction}
Due to recent significant advancements in fabricating low-loss optical,
electrical and mechanical devices, we will soon be able to probe behaviors
of macroscopic mechanical oscillators in the quantum regime. This will not
only shed light on quantum-limited measurements of various physical quantities,
such as a weak force, but also help us to achieve a better understanding of
quantum mechanics on macroscopic scales.

As a premise of investigating macroscopic quantum mechanics (MQM), the mechanical
oscillator should be prepared close to be in a pure quantum state. To achieve this,
there are mainly three approaches raised in the literature: (i) The first and the
most transparent approach is to cool down the oscillator by coupling it to an additional
heat bath that has a temperature $T_{\rm add}$ much lower than that of the environment
$T_{0}$. As a result, the oscillator will achieve an effective temperature given by $T_{\rm eff}=(T_0 \,\gamma_m+T_{\rm add}\,\Gamma_{\rm add})/(\gamma_m+\Gamma_{\rm add})$ with $\gamma_m$
and $\Gamma_{\rm add}$ denoting the damping due to coupling to the environment and
the additional heat bath, respectively. In the strong-damping regime with
$\Gamma_{\rm add}\gg \gamma_m$, we achieve the desired outcome with $T_{\rm eff}\approx T_{\rm add}$.
Since the typical optical frequency $\omega_0$ can be much higher than $k_B T_0/\hbar$, a
coherent optical field can be effectively served as a
zero-temperature heat bath. Indeed, by coupling an oscillator parametrically
to an optical cavity, many state-of-the-art experiments have demonstrated significant
cooling of the oscillator, achieving a very low thermal occupation number \cite{Metzger, Gigan, Arcizet,
Kleckner, Schliesser1, Corbitt1, Corbitt2, Schliesser2, Favero, Teufel, Thompson,
Lowry, Groblacher, Schediwy, Jourdan, Aspelmeyer}. Similar mechanism also applies to
the electromechanical system as demonstrated in the experiments \cite{Blair, Naik, Schwab};
(ii) The second approach is to introduce additional damping via feedback, i.e., the so-called
cold-damping. The feedback loop modifies the dynamics of the oscillator in a way similar
to the previous cooling case. Such an approach has also been realized experimentally
\cite{Cohadon,Poggio,LIGO}. If the intrinsic mechanical and eletrical/optical
qualities of the coupled system are high, those cooling and cold-damping experiments
can eventually achieve the quantum ground state
of a mechanical oscillator \cite{Vyatchanin_cooling, Mancini, Marquardt, Rae, Genes, ctrl};
(iii) The third approach is to construct a conditional quantum state of the mechanical
oscillator via continuous position measurements. Quantum mechanically, if the oscillator
position is being continuously monitored, a certain classical trajectory in the phase
space can be mapped out, and the oscillator is projected into {\it a posteriori state}
\cite{Barchielli1992} which is also called a conditional quantum state~\cite{Mil1996,Gar2004,
DTPW1999, HJHS2003,MRSDC2007,state_pre}. Given an ideal continuous measurement without loss,
the resulting conditional quantum state of the oscillator is a pure state.

Recently, we theoretically investigated the third approach for general linear position
measurements in great details ~\cite{state_pre}. The analysis of this work is independent
of the scale and mass of the oscillator --- these parameters will only modify the structure
of arising noises. In particular, we applied our formalism to discuss MQM experiments
with macroscopic test masses in future gravitational-wave (GW) detectors. We demonstrated
explicitly that given the noise budget for the design sensitivity, next-generation GW detectors
such as Advanced LIGO \cite{AdvLIGO} and Cryogenic Laser Interferometer Observatory (CLIO)
\cite{CLIO} can prepare nearly pure Gaussian quantum states and create Einstein-Podolsky-Rosen-type
entanglement between macroscopic test masses. Besides, we showed that the free-mass
{\it Standard Quantum Limit} (SQL) ~\cite{SQL1,SQL2,SQL3} for the detection sensitivity
\begin{equation} \label{Eq:SQL}
S_x^{\rm SQL}(\Omega) =\frac{2\hbar}{m\, \Omega^2}\,,
\end{equation}
where $m$ is mass of the probing test mass and $\Omega$ is the detection frequency, also
serves as a benchmark for MQM experiments with GW detectors.

\begin{figure}
\includegraphics[width=0.35\textwidth, bb=0 0 409 212,clip]{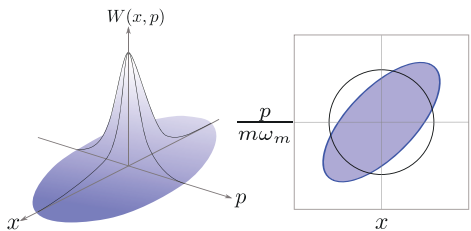}
\caption{(Color online) A schematic plot of a Wigner function $W(x,p)$ (left) and the
corresponding uncertainty ellipse for the covariance matrix ${\bf V}^{\rm cond}$
(can be viewed as a projection of the Wigner function). The center of the plot is given
by the conditional mean $(x^{\rm cond}, p^{\rm cond})$. The Heisenberg limit is shown
in a unit circle with radius given by the zero-point fluctuation $\hbar/(2m\omega_m)$.
For a pure Gaussian
conditional quantum state, the area of the ellipse, i.e., $\pi\det {\bf V}^{\rm cond}/(2m\omega_m)^2$,
is also equal to that of Heisenberg limit. Therefore, the uncertainty product
$\det {\bf V}^{\rm cond}$ can be served as an appropriate figure of merit for
quantifying purity of a quantum state. \label{wf}}
\end{figure}

More concretely, a Gaussian conditional quantum state is fully described by its Wigner
function as shown schematically in Fig. \ref{wf}. It is given by
\be\label{wigf}
W(x, p)=\frac{1}{2\pi \sqrt{\det {\bf V}^{\rm cond}}} \exp\left[-\frac{1}{2}
{\vec X}\,{{\bf V}^{\rm cond}}^{-1}\,{\vec X}^T\right].
\ee
Here ${\vec X}=[x-x^{\rm cond}, p-p^{\rm cond}]$ with $x^{\rm cond}$ and $p^{\rm cond}$
denoting conditional means of oscillator position $x$ and momentum $p$, and
${\bf V}^{\rm cond}$ is the covariance matrix between position and momentum.
Purity of the conditional quantum state can be quantified by the uncertainty product
which is defined as
\be\label{purity}
U\equiv\frac{2}{\hbar}\sqrt{\det {\bf V}^{\rm cond}}=\frac{2}{\hbar}\sqrt{V_{xx}^{\rm cond}V_{pp}^{\rm cond}-{V_{xp}^{\rm cond}}^2}
\ee
which is also proportional to square root of the area of the uncertainty ellipse
as shown in Fig. \ref{wf}. In Ref. \cite {state_pre}, we related this uncertainty
product $U$ of the conditional quautum state of test masses in GW detectors
to the SQL-beating ratio of the classical noise, and the amount of entanglement
between test masses to the size of the frequency window (ratio between upper and
lower ends of that frequency window) in which the classical noise goes below the SQL.

A state-preparation stage alone does not provide a complete test of MQM.
This is because the measurement data in the state-preparation process only
allow us to measure a classical trajectory of the oscillator -- quantum fluctuations
are only inferred from the noise budget, but not directly visible. Therefore, the
resulting conditional quantum state critically relies on the noise model of the
measurement device. If such noise model is imprecise, it will yield severe
discrepancies between the actual quantum state and the conditional one. Therefore,
it calls for a second measurement stage which has to follow up the preparation stage.
In this paper, we will address the above issue by considering a subsequent
state-verification procedure, in which we make a tomography of the conditional quantum state
obtained during the preparation stage. On the one hand, this verification stage can
serve as a check of the specific noise model to verify the prepared quantum state.
On the other hand, if we insert an evolution stage with the oscillator evolving
freely before the verification, the quantum dynamics of the oscillator can also be
probed, which allows us to study different decoherence effects and also check
whether a macroscopic mechanical oscillator does evolve in the same way as a
quantum harmonic oscillator or not.

\begin{figure}
\includegraphics[width=0.184\textwidth, bb=0 0 274 274,clip]{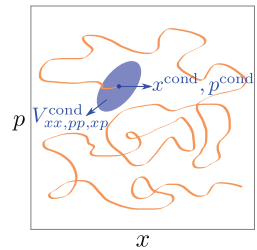}
\caption{(Color online) A schematic plot of random walk of the conditional quantum
state, i.e., its Wigner function, in the phase space. Its center is given by the
conditional mean $[x^{\rm cond}(t),\,p^{\rm cond}(t)]$, with the uncertainty given by
conditional variances $V_{xx, pp, xp}^{\rm cond}$. To verify the prepared conditional quantum
state, the only knowledge that verifier needs to know is classical information
of the conditional mean provided by the preparer if noises are Markovian. \label{rw}}
\end{figure}

Since the conditional quantum state undergoes a random walk in the phase space as
shown schematically in Fig. \ref{rw}, classical information of the conditional
mean, obtained by the preparer from the measurement data, needs to
be passed onto the verifier who will then proceed with a tomography process. Suppose
the state preparation stage ends at $t=0$ and the preparer obtain a conditional quantum whose
Wigner function is $W(x(0), p(0))$. The task of the verifier is trying to reconstruct this
Wigner function by synthesizing marginal
distributions of different mechanical quadratures $\hat X_{\zeta}(0)$ from ensemble
measurements at $t>0$, and
\begin{equation}\label{ROquad}
\hat X_{\zeta}(0)\equiv  \hat x(0)\cos\zeta+\frac{
\hat p(0)}{m\,\omega_m}\sin\zeta,
\end{equation}
where $\hat x(0)$ and $\hat p(0)$ denote oscillator position and momentum
at $t=0$ and $\omega_m$ is the oscillation frequency. This process is similar to the
optical quantum tomography where different optical quadratures are measured with
homodyne detections \cite{Lvovsky}. However, there is one significant difference ---
mechanical quadratures are not directly accessible with linear position measurements
which measure
\be\label{position}
\hat x_q(t)=\hat x(0)\cos\omega_m t+\frac{\hat p(0)}{m\,\omega_m}\sin \omega_m t,
\ee
rather than $\hat X_{\zeta}$. To probe mechanical quadratures, we propose the use
a time-dependent homodyne detection with the local-oscillator phase varying in time. Given a
measurement duration of $T_{\rm int}$, we can construct an integral estimator, which reads
\be\label{intest}
\hat X=\int_{0}^{T_{\rm int}}dt\,g(t)\,\hat x(t)\propto\hat x(0)\cos\zeta'+\frac{\hat p(0)}
{m\,\omega_m}\sin\zeta'
\ee
with $\cos\zeta'\equiv\int_{0}^{T_{\rm int}}dt\,g(t)\cos\omega_m t$ and $\sin\zeta'\equiv\int_{0}^{T_{\rm int}}dt\,g(t)\sin \omega_m t$.
Therefore, a mechanical quadrature $\hat X_{\zeta'}$ is probed [cf. Eq. \eqref{ROquad}].
Here $g(t)$ is some filtering function, and it is determined by the time-dependent homodyne phase and also the way how data at different time are combined.

\begin{figure}
\includegraphics[width=0.43\textwidth, bb=0 0 476 187,clip]{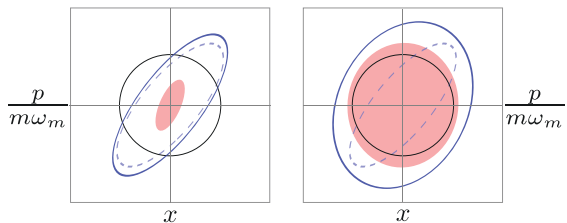}
\caption{(Color online) A schematic plot of the uncertainty ellipses of reconstructed states
with the same prepared Gaussian quantum state but different levels of verification accuracy,
which shows the necessity of a sub-Heisenberg accuracy. The center of the plot is given by
the conditional mean $(x^{\rm cond}, p^{\rm cond})$.  The shaded regimes correspond to the
verification accuracy.  The Heisenberg limit is shown by a unit circle. The dashed and solid ellipses
represent the prepared state and the reconstructed states respectively.
\label{ad}}
\end{figure}

The ability to measure mechanical quadratures does not guarantee success of a verification
process. In order to recover the prepared quantum state, it requires a verification accuracy
below the Heisenberg limit. Physically, the output of the verification process is a sum
of the mechanical-quadrature signal and some uncorrelated Gaussian noise. Mathematically,
it is equivalent to applying a Gaussian filter onto the original Wigner function $W(x,p)$
of the prepared state \cite{Yurke}, and thus the reconstructed Wigner function
is
\be\label{Wre}
W_{\rm recon}(x, p)=\int_{-\infty}^{\infty}dx'dp'\,\psi(x-x', p-p') W(x',p')
\ee
where the Gaussian filter $\psi(x,p)$ is given by
\be\label{vadef}
\psi(x, p)\equiv\frac{1}{2\pi \sqrt{\det {\bf V}^{\rm add}}} \exp\left[-\frac{1}{2}{\vec \xi}\,{{\bf V}^{\rm add}}^{-1}{\vec \xi }^T\right]
\ee
with $\vec \xi=[x, p]$ and ${\bf V}^{\rm add}$ denoting the covariance matrix for
the added verification noise. If the prepared quantum state is Gaussian, using the property
of Gaussian integration, the reconstructed Wigner function reads
\be
W_{\rm recon}(x, p)=\frac{1}{2\pi \sqrt{\det {\bf V}^{\rm recon}}}
\exp\left[-\frac{1}{2}{\vec \xi}\,{{\bf V}^{\rm recon}}^{-1}{\vec \xi }^T\right],
\ee
and the covariance matrix ${{\bf V}^{\rm recon}}$ is
\be\label{Vplus}
{\bf V}^{\rm recon}={\bf V}^{\rm cond}+{\bf V}^{\rm add}.
\ee
In Fig. \ref{ad}, we show schematically the effects of different levels of verification
accuracy given the same prepared conditional quantum state. A sub-Heisenberg accuracy,
with an error area smaller than the Heisenberg limit, is
essential for us to obtain a less distorted understanding of the original prepared quantum state.
In addition, if the prepared quantum state of the mechanical oscillator is non-Gaussian
\cite{Mancini_nonclassical, Bose, Lifshiz, Jacobs, Khalili}, a sub-Heisenberg accuracy is
a necessary condition for unveiling the non-classicality of the quantum state as shown
schematically in Fig. \ref{negw} and proved rigorously in the Appendix \ref{App0}.

Verifications of quantum states below the Heisenberg limit also naturally allow us to
test whether entanglement between two macroscopic test masses in GW detectors can indeed be established,
as predicted in Ref.~\cite{MRSDC2007,state_pre}, and how long such entangled state
can survive. Survival of macroscopic entanglement can test one particular
version of gravity decoherence conjectured by Di\'{o}si~\cite{Diosi} and Penrose~\cite{Penrose}.
For an individual object, it is not entirely clear the classical superposition
of what {\it pointer} states gravity decoherence will drive it into. For an
entangled state among multiple objects, even though Gaussian, it would naturally
have to decay into the one that is not entangled, within the gravity decoherence timescale.

\begin{figure}
\includegraphics[width=0.32\textwidth, bb=0 0 500 264,clip]{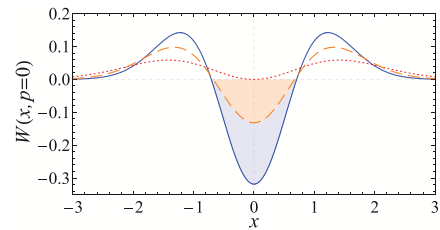}
\caption{(Color online) Values of reconstructed Wigner functions on the $p = 0$ plane,
i.e., $W_{\rm recon}(x, p = 0)$, for a single-quantum state, obtained at
different levels of verification accuracy. Solid curve shows the ideal case with
no verification error. Dashed and dotted curves correspond to the cases with a
verification error of $1/4$ and $1/2$ of the Heisenberg limit, respectively. The
negative regime (shaded) or the non-classicality vanishes as the verification error
increases. This again manifests the importance of a sub-Heisenberg verification
accuracy.\label{negw}}
\end{figure}

As we will show, in order to achieve a sub-Heisenberg accuracy, we need to optimize
the local-oscillator phase of the time-dependent homodyne detection as well as the
weight with which data collected at different time will be combined. If there is no
readout loss, this optimization automatically will give a detection scheme that evades measurement-induced
back action, the same as the variational-type measurement scheme
proposed by Vyatchanin and Matsko ~\cite{bae} for detecting gravitational-wave
signals with known arrival time. Since in a single
measurement setup, different quadratures do not commute with each other, namely
\be
[\hat X_{\zeta},\,\hat X_{\zeta'}]=\frac{i\,\hbar}{m\,\omega_m}\sin(\zeta-\zeta'),
\ee
one needs multiple setups and each makes ensemble measurements of one particular
quadrature $\hat X_{\zeta}$ with a sub-Heisenberg accuracy --  the synthesis of
these measurements yields a quantum tomography.

As a sequence to Ref. \cite{state_pre} and to motivate MQM experiments with future
GW detectors, we will also focus on the same parameter regime where the characteristic
measurement frequency is much higher than the oscillator frequency and the oscillator
can be treated as a free mass. In addition, we will consider situations
where the spectra of the classical noise can be modeled as being white.
Non-Markovianity of noise sources -- although they certainly
arise in actual GW detectors~\cite{state_pre} and will be crucial for the success of
a real experiment --- is a rather technical issue. The non-Markovianity will not change
the results presented here significantly, as we will show and address in a separate paper
\cite{chen1}.

This paper is organized as follows: in Sec. \ref{Model}, we will formulate the system
model mathematically by writing down the Heisenberg equations of motion; in Sec.~\ref{order},
we will provide a timeline for a full MQM experiment with preparation, evolution and
verification stages, and use simple order-of-magnitude estimates to show
that this experimental proposal is indeed plausible; in Sec.~\ref{markovian}, we will evaluate the
verification accuracy in the presence of Markovian noises (largely confirming the order-of-magnitude
estimates, but with precise numerical factors); in Sec.~\ref{ent}, we will consider verifications of
macroscopic quantum entanglement between test masses in GW detectors as a test
of gravity decoherece; in Sec.~\ref{con}, we will summarize our main results. In the Appendix,
we will present mathematical details for solving integral equations that we encounter in obtaining
the optimal verification scheme.

\section{Model and Equations of Motion}\label{Model}

\begin{figure*}
\includegraphics[width=0.85\textwidth, bb= 0 0 594 224,clip]{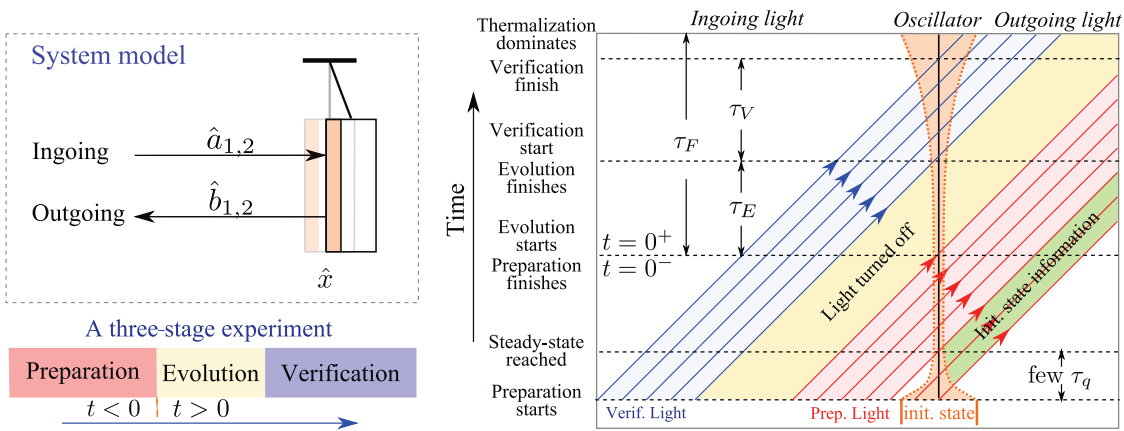}
\caption{A schematic plot of the system (upper left panel) and the corresponding spacetime diagram
(right panel) showing the timeline of the proposed MQM experiment (see Sec.~\ref{subsec:timeline}
for detailed explanations). In this schematic plot, the oscillator position is denoted by $\hat x$ which
is coupled to the optical fields through radiation pressure. The ingoing and outgoing optical fields
are denoted by $\hat a_{1, 2}$ and $\hat b_{1,2}$ with subscripts $1, 2$ for the amplitude and phase
quadratures, respectively. In the spacetime diagram, the world line of the oscillator is shown by the
middle vertical line. For clarity, ingoing and outgoing optical fields are represented
by the left and right regions on the different sides of the oscillator world line, even though in reality,
optical fields escape from the same side as where they enter. We show light rays during preparation and verification stages in red and blue. In between, the yellow shaded region describes the evolution stage with light turned off for a duration of $\tau_E$.
The conditional variance of the oscillator motion is represented by the shaded region alongside the
central vertical line (not drawn to the same scale as light propagation). At the beginning of
preparation, the conditional variance is dominated by that of the initial state (orange).
After a transient, it is determined by incoming radiations and measurements.
Right after state preparation, we show the expected growth of the conditional variance due to thermal
noise alone, and ignoring the effect of back-action noise which is evaded during the
verification process. The verification stage lasts for a duration of
$\tau_V$, and it is shorter than $\tau_F$ after which the oscillator will be dominated by thermalization. \label{ST}}
\end{figure*}

In this section, we will present a mathematical description of the system model, as
shown schematically in the upper left panel of Fig. \ref{ST}. The oscillator
position is linearly coupled to coherent optical fields through radiation pressure. Meanwhile,
information of the oscillator position flows into the outgoing optical fields continuously.
This models a measurement process in an optomechanical system without a cavity or with a
large-bandwidth cavity. The corresponding Heisenberg equations, valid for both preparation
and verification stages, are formally identical to classical equations of motion except for
that all quantities are Heisenberg operators. The oscillator position $\hat x$
and momentum $\hat p$ satisfy the following equations:
\begin{align} \label{11}
\dot{\hat x}(t)&=\hat p(t)/m,\\\label{12}
\dot {\hat p}(t)&=-2\gamma_m {\hat p}(t)-m\,\omega_m^2
{\hat x}(t)+ \alpha\,\hat a_1 (t) + \hat\xi_F (t).
\end{align}
Here $\alpha\, \hat a_1$ corresponds to the quantum-radiation-pressure noise or so-called
back-action noise; $\alpha \equiv (\hbar\, m \,\Omega_q^2)^{1/2} = (8\,I_0\,\omega_0\, \hbar/c^2)^{1/2}$
is the coupling constant between the oscillator and optical fields with $I_0$ denoting the optical
power and $\Omega_q$ quantifying the characteristic frequency of measurement strength.
We have included the fluctuation-dissipation mechanism of the mechanical oscillator
by introducing the mechanical damping rate $\gamma_m$ and classical-force noise $\hat \xi_{F}$, i.e.,
the Brownian thermal noise. In the Markovian limit, the correlation function for $\hat \xi_{F}$ is given by \footnote{\label{ena} Here $\langle\,\;\rangle_{\rm sym}$ stands for a \textit{symmetrized} ensemble
average. For a system characterized
by a density matrix $\hat \rho$, it is defined as
$$\langle \hat o_1(t)\,\hat o_2(t')\rangle_{\rm sym}\equiv
\mathrm{Tr}\left\{[\hat o_1(t)\hat o_2(t')+\hat o_2(t')\hat
o_1(t)]\hat \rho\right\}/2\,.$$}
\be\label{Fthc}
\langle\hat \xi_{F}(t)\,\hat \xi_{F}(t') \rangle_{\rm sym}=S_F^{\rm th} \delta (t-t')/2
\ee
where $S_F^{\rm th}= 4m\gamma_m k_B T_0\equiv 2\hbar\,m\,\Omega_F^2$ and we have defined a
characteristic frequency $\Omega_F$ for the thermal noise.

The amplitude and phase quadratures of ingoing optical fields $\hat a_{1,2}$ and of outgoing optical fields $\hat b_{1,2}$ satisfy the following input-output relations:
\begin{align}\label{13}
\hat b_1 (t) &=\sqrt{\eta}\,\hat n_1 (t)+\sqrt{1-\eta}\,\hat a_1 (t),\\\label{14}
\hat b_2 (t) &=\sqrt{\eta}\,\hat n_2 (t)+\sqrt{1-\eta}\left[\hat a_2 (t) +
\frac{\alpha}{\hbar}\hat x (t) + \frac{\alpha}{\hbar} \hat \xi_{x} (t)\right].
\end{align}
Here $\hat n_{1,2}$ originate from nonunity quantum efficiency
of the photodetector for $\eta>0$. In the paraxial and narrow-band approximation, $\hat a_{1,2}$ are related to the electrical-field strength at the central frequency $\omega_0$ by
\cite{Loudon, Caves, klmtv}:
\be
\hat
E(t)\equiv \left(\frac{{4\pi\hbar\,\omega_0}}{{\mathcal{S}\,c}}\right)^{1/2} \{[\bar a+\hat a_1(t)]
\,\cos\omega_0 t+\hat a_2(t)\,\sin\omega_0 t\}
\ee
with $\bar a$ denoting the classical amplitude and $\mathcal{S}$ standing for the effective
cross-section area of the laser beam. A similar relation also holds for the
outgoing fields $\hat b_{1,2}$. In addition, they satisfy
$[\hat a_1(t), \,\hat a_2(t')]=[\hat b_1(t), \,\hat b_2(t')]=i\,\delta(t-t')$. Their
correlation functions read
\be
\langle \hat a_i(t)\,\hat a_j(t')\rangle_{\rm sym}=\delta_{ij}e^{\pm 2q}\delta(t-t')/2, \;(i,j=1,2)
\ee
where $q$ denotes the squeezing factor ($q=0$ for a vacuum-state input) with ``$+$" for
the amplitude quadrature and ``$-$" for the phase quadrature. Correspondingly, the correlation function for the back-action noise
$\alpha\, \hat a_1$ is simply
\be\label{FBAc}
\langle \alpha\,\hat a_1(t)\,\alpha\,\hat a_1(t')\rangle_{\rm sym}=S_F^{\rm BA} \delta(t-t')/2
\ee
with $S_F^{\rm BA}\equiv e^{2q}\hbar\,m\,\Omega_q^2$. In Eq. \eqref{14}, $\hat \xi_x$ is the
sensing noise. One example is the internal thermal noise, and it is defined as the difference between 
the center of mass motion and the surface motion of the oscillator which is actually being measured.
In the Markovian approximation, it has the following correlation function:
\be\label{Sxthc}
\langle\hat \xi_{x}(t)\,\hat \xi_{x}(t') \rangle_{\rm sym}=S^{\rm th}_x\delta(t-t')/2
\ee
where $S^{\rm th}_x= {\hbar}/({m\,\Omega_x^2})$ and we introduce a characteristic
frequency $\Omega_x$ for the sensing noise.

Note that the $\Omega_q$, $\Omega_F$, and $\Omega_x$ that we have introduced are also the frequencies at which the
back-action noise, thermal noise and sensing noise intersect the SQL [cf. Eq.~\eqref{Eq:SQL}],
respectively. They are identical to what were introduced in Ref. \cite{state_pre}. 
For conveniences of later discussions, we introduce the following
dimensionless ratios:
\be
\zeta_F=\Omega_F/\Omega_q, \quad \zeta_x =\Omega_q/\Omega_x.
\ee
In addition, we define two characteristic timescales for the measurement and thermal-noise strength as
\begin{equation}\label{ts}
\tau_q \equiv 1/\Omega_q \,,\quad \tau_F\equiv 1/\Omega_F\,.
\end{equation}

\section{Outline of the experiment with order-of-magnitude estimate}
\label{order}
In this section, we will describe in details the timeline of a plausible MQM experiment
(subsection \ref{subsec:timeline}) and provide order-of-magnitude estimates of the conditional
variance of the prepared quantum state, the evolution of the prepared quantum state,
and the verification accuracy in the free-mass regime (subsections \ref{preparation}, \ref{subsec:evo} and \ref{subsec:verify}). This will provide qualitatively the requirements on the noise level for the success of a MQM experiment. We will give more rigorous treatments in the next section.

\subsection{Timeline of proposed experiment}
\label{subsec:timeline}

We have sketched a space-time diagram for the proposed MQM experiment in the right panel of Fig.~\ref{ST}
--- with time going upward, therefore we start from the bottom of the figure.

\noindent {\B{\it Lock Acquisition.}} At the beginning, the mechanical oscillator
is in a highly mixed state, so are the optical fields. Therefore, the first step is
to ``acquire lock'' of the measurement device, and reach a steady-state operation mode,
during which several $\tau_q$ will have elapsed. From this time and on, initial-state
information will have been forgotten (propagating outward within the green
strip), and the state of the oscillator will be determined by the driving fields, including
the classical-force noise and sensing noise, as well as the quantum noise. This will
be the start of the state-preparation stage (region above the $45^\circ$ green strip).

\noindent{\B{\it State Preparation.}} This stage is a steady-state operation of the
measurement device. The quantum state of the oscillator is collapsed continuously due
to homodyne readouts of the photocurrent. At any instant during state preparation, based
on the measured history of the photocurrent (mostly on data within several times $\tau_q$ to
the past of $t$), the conditional expectation $(x^{\rm cond},\,p^{\rm cond})$ for the
oscillator position $\hat x$ and momentum $\hat p$ can be constructed. The second moments, describable
by the covariance matrix between position and momentum, which consists of $V_{xx}^{\rm
cond}$, $V_{xp}^{\rm cond}$ and $V_{pp}^{\rm cond}$, can be calculated from
the noise model of the measurement device --- they, together with $x^{\rm cond}$ and
$p^{\rm cond}$, fully determine the quantum state, i.e., the Wigner function of the
oscillator at any instant [cf. Eq. \eqref{wigf}]. For a Gaussian steady state, the
construction of $(x^{\rm cond},\,p^{\rm cond})$ and conditional covariance matrix from
the history of the photocurrent can be accomplished most easily using Wiener Filtering,
as shown in Ref.~\cite{state_pre}.

The preparation stage terminates at $t=0$, when $(x^{\rm cond},\,p^{\rm cond})$
and covariance matrix will be determined by data from several $-\tau_q$ up to $0$ as
shown by the red strip.

\noindent{\B{\it State Evolution.}} If we want to investigate the quantum dynamics
of the oscillator and study various decoherence effects, we can delay the verification
process and allow the oscillator to freely evolve with the interaction light turned off
(represented by the yellow strip). During this period, the thermal noise will induce
diffusions of the oscillator position and momentum, thus increasing the conditional variance
as shown schematically by broadening of the shaded
region alongside the oscillator world line. If there were any additional decoherence effect,
the variance will grow faster than the case with the thermal decoherence alone. A follow-up
verification allows us to check whether additional decoherence mechanisms, such as the gravity
decoherence conjectured by Di\'{o}si \cite{Diosi} and Penrose \cite{Penrose}, exist or not.

\noindent{\B{\it State Verification.}} After the evolution stage, the verification
stage starts (represented by blue strip). We intentionally use different colors to
label the preparation light and verification light --- symbolizing the fact that
in principle, a different observer (verifier) could perform the verification process, and
verify the quantum state by him/herself. The only knowledge from the preparer would
be the conditional expectation $x^{\rm cond}$ and $p^{\rm cond}$ if all noise sources
are Markovian. The verifier uses a time-dependent homodyne detection and collects the data from
measuring the photocurrents. The verification process lasts for a timescale of $\tau_V$ between 
the characteristic measurement timescale $\tau_q$ and the thermal
decoherence timescale $\tau_F$, after which diffusions of $\hat x$ and $\hat p$ in the
phase space become much larger than the Heisenberg limit. Based upon the measurement data,
the verifier can construct an integral estimator for one particular mechanical quadrature [cf. Eq. \ref{intest}].

The above three stages have to be repeated for many times before enough data
are collected to build up statistics. After finishing the experiment, the verifier will
obtain a reconstructed quantum state of the mechanical oscillator, and then can
proceed to compare with the preparer and interpret the results.

\subsection{Order-of-magnitude estimate of the conditional variance} \label{preparation}
In this and the following two subsections, we will provide order-of-magnitude estimates for
a three-staged MQM experiment including preparation, evolution and verification stages. This
gives us physical insights into different timescales involved in a MQM experiment and also
the qualitative requirements for an experimental realization. We will justify those estimates
based upon more careful treatments in the next several sections.

Based upon the measurement data from several $-\tau_q$ to 0, one can construct a conditional
quantum state for the mechanical oscillator. Suppose that the phase quadrature
of the outgoing fields is being measured and the photodetection is ideal with $\eta=0$.
Given a measurement timescale of $\tau$ (measuring from $-\tau$ to $0$), variances for the oscillator position and momentum at $t=0$ in the free-mass regime with $\omega_m\rightarrow 0$ are approximately equal to [cf. Eqs. \eqref{11}, \eqref{12}, \eqref{13} and \eqref{14}]
\begin{eqnarray}\label{25}
\delta x^2(0) \!\!\!&\sim& \!\!\! S_x^{\rm tot}/{\tau} +
{\tau^3 S_F^{\rm tot}}/{m^2} \sim N_x^{\frac{3}{4}} N_F^{\frac{1}{4}} \delta x_q^2,\\\label{26}
\delta p^2(0) \!\!\!&\sim&\!\!\! {m^2 S_x^{\rm tot}}/{\tau^3} + {\tau
S_F^{\rm tot}} \sim N_x^{\frac{1}{4}}N_F^{\frac{3}{4}} \delta
p_q^2.
\end{eqnarray}
Here $S_F^{\rm tot}\equiv S_F^{\rm BA} + S_F^{\rm th}$ [cf. Eqs. \eqref{Fthc} and \eqref{FBAc} ] and
$S_x^{\rm tot}\equiv S_x^{\rm sh} + S_x^{\rm th}$ with $S_x^{\rm sh}$ denoting
the shot noise due to $\hat a_2$ [cf. Eqs. \eqref{14} and \eqref{Sxthc}]; we have defined
\begin{equation}
N_x \equiv  1+2\,\zeta_x^2 \,,\quad N_F \equiv 1+2\,\zeta_F^2,
\end{equation}
while
\begin{equation}\label{dxq}
\delta x_q^2 \equiv {\hbar}/({2m\,\Omega_q})\,,\quad \delta p_q^2
\equiv  {\hbar \,m\,\Omega_q}/2.
\end{equation}
The optimal measurement timescale is given
by $\tau \sim\tau_q$. Purity of the prepared conditional quantum
state at $t=0$ is approximately equal to  [cf. Eq. \eqref{purity}]
\be\label{pesti}
U(0)\sim \frac{2}{\hbar}\delta x(0)\,\delta p(0)\sim N_x N_F.
\ee
If classical noises are low, namely, $N_x\sim N_F\sim 1$, the conditional quantum
state will be pure with $U(0)\sim1$. For future GW detectors such as AdvLIGO, both
$\zeta_x$ and $\zeta_F$ will be around $0.1$, and such a low classical-noise
budget clearly allows us to prepare nearly pure quantum states of the macroscopic test masses.

\subsection{Order-of-magnitude estimate of state evolution}
\label{subsec:evo}
\begin{figure}
\includegraphics[width=0.25\textwidth, bb=0 0 232 232,clip]{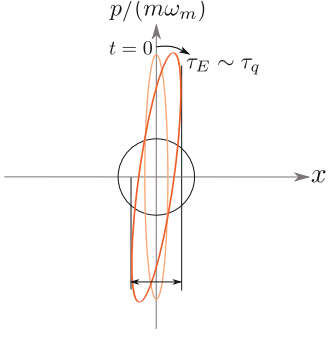}
\caption{ (Color online) Rotation and diffusion of a highly position-squeezed conditional quantum state prepared by a strong measurement with $\Omega_q\gg\omega_m$. The initial-momentum uncertainty will
contribute an uncertainty in the position comparable to the initial position uncertainty
when the evolution duration $\tau_E\sim \tau_q$. \label{squeeze}}
\end{figure}

During the evolution stage, the uncertainty ellipse of the conditional quantum state
will rotate at the mechanical frequency in the phase space, and meanwhile there is a growth
in the uncertainty due to
thermal decoherence as shown schematically in Fig. \ref{squeeze}. Given a strong measurement,
the variance of the resulting conditional quantum state in position $\delta x^2(0)$ will be approximately equal to $\delta x_q^2$ as shown explicitly in Eq. \eqref{25} with $N_x, N_F\sim 1$. It is much smaller than the zero-point uncertainty of an $\omega_m$ oscillator, which is given by $\hbar/(2 m\,\omega_m)$. Therefore, the conditional quantum state of the oscillator is highly squeezed in position. The position uncertainty contributed by the initial-momentum will be comparable to that of the initial-position uncertainty after a evolution duration of $\tau_q$. This can be directly seen from an order-of-magnitude estimate. In the free-mass regime,
\be
\hat x(t)\sim \hat x(0)+\frac{\hat p(0)}{m} t.
\ee
For an evolution duration of $\tau_E$, the corresponding variance in position is
\be\label{29}
\delta x^2(\tau_E)\sim \delta x^2(0)+\frac{\delta p^2(0)}{m^2}\tau_E^2\sim \delta x^2(0)[1+(\Omega_q \tau_E)^2].
\ee
The contribution from the initial-momentum uncertainty (the second term) will become important when $\Omega_q\tau_E\sim 1$ or equivalently $\tau_E\sim \tau_q$.

Apart from a rotation, the uncertainty ellipse will also grow due to thermal decoherence. Variances in the position and momentum contributed by thermal decoherence are approximately given by [cf. Eqs. \eqref{11} and \eqref{12}]
\begin{align}\label{30}
\delta x_{\rm th}^2(\tau_E)&\sim \tau_E^3S_F^{\rm th}/m^2= \zeta_F^2(\Omega_q\tau_E)^3\delta x_q^2,\\
\delta p_{\rm th}^2(\tau_E)&\sim \tau_E\,S_F^{\rm th}=\zeta_F^2(\Omega_q \tau_E)\delta p_q^2.\label{31}
\end{align}
The growth in the uncertainty ellipse will simply be
\be\label{33}
U^{\rm th}(\tau_E)\sim \frac{2}{\hbar}\delta x_{\rm th}(\tau_E)\delta p_{\rm th}(\tau_E)\sim \zeta_F^2(\Omega_q\tau_E)^2=(\tau_E/\tau_F)^2.
\ee
When $\tau_E>\tau_F$, $U^{\rm th}(\tau_E)>1$ and the conditional quantum state will be dominated by thermalization.

If there were any additional decoherence effect, the growth in the uncertainty will be much larger
than what has been estimated here. A subsequent verification stage can serve as a check.

\subsection{Order-of-magnitude estimate of the verification accuracy}
\label{subsec:verify} To verify the prepared conditional quantum state,
 the oscillator position needs to be measured for a finite duration to obtain information about
$\hat x(0)$ and $\hat p(0)$ [cf. Eq. \eqref{position} and \eqref{intest}] or about $\hat x(\tau_E)$ and
$\hat p(\tau_E)$ if the evolution stage is inserted.
In order for an entire state characterization to be possible, one might then expect
that an oscillation period must pass, and during this period, the thermal noise should
cause an insignificant diffusion of the oscillator momentum compared with its zero-point
uncertainty, which requires \cite{SQL3}
\be
\frac{k_B T_0}{\hbar\,\omega_m}<Q_m
\ee
with $Q_m\equiv \omega_m/(2\gamma_m)$ denoting the mechanical quality factor. This requirement
is unnecessary if the initial quantum state is prepared by a strong measurement. As we have mentioned
in the previous subsection, the resulting condition quantum state is highly squeezed in position and the
initial-momentum uncertainty will make a significant contribution to the uncertainty in position after
$\tau>\tau_q$. This means, depending on the particular strategy, one can
extract $\hat x$ and $\hat p$ below the levels of $\delta x_q$ and $\delta p_q$, respectively, as long as one is able to measure oscillator position with an accuracy better than $\delta x_q$, within a timescale of several $\tau_q$. This is certainly possible if
the measurement-induced back action is evaded.

To evade the measurement-induced back action, one notices the fact that
the amplitude quadrature $\hat b_1$ contains $\hat a_1$ which
is responsible for the back action, and meanwhile the phase quadrature $\hat b_2$ contains the information of oscillator position, part of which is contributed by the back action [cf. Eq. \eqref{11}-\eqref{14}].
Therefore, if we measure particular combinations of $\hat b_1$ and $\hat b_2$ at different times,
by summing up those measurements, we will be able to cancel the back action and obtain
a back-action-evading (BAE) estimator for a given mechanical quadrature. Such cancelation mechanism is only limited by the readout loss ($\eta\neq0$), which introduces uncorrelated vacuum fluctuations.

We can make an order-of-magnitude estimate to show that a sub-Heisenberg accuracy can be indeed achieved. With the BAE technique, the force noise that limits the verification
accuracy will only contain the thermal-noise part. Similar to Eqs. \eqref{25} and \eqref{26} but with $S_F^{\rm tot}$ replaced by $S_F^{\rm th}$, the variances in position and momentum during the verification stage are simply
\begin{eqnarray}
\delta x_V^2 \!\!&\sim&\!\! {S_x^{\rm tot}}/{\tau} +{\tau^3
S_F^{\rm th} }/{m^2}
\sim N_x^{3/4}\zeta_F^{1/2} \delta x_q^2\,, \\
\delta p_V^2  \!\! &\sim&\!\! m^2 S_x^{\rm tot}/{\tau^3} + \tau
S_F^{\rm th} \sim N_x^{1/4} \zeta_F^{3/2} \delta p_q^2\,.
\end{eqnarray}
Here the optimal verification timescale would be $\tau_V \sim \zeta_F^{-1/2}\tau_q $ and $\tau_q< \tau_V <\tau_F$. A summarizing figure of merit for the verification accuracy
is approximately given by
\begin{equation}
U^{\rm add}|_{\rm BAE}\sim \frac{2}{\hbar}\delta x_V \delta p_V \sim N_x^{1/2} \zeta_F.
\end{equation}
A sub-Heisenberg accuracy can be achieved when $\zeta_F <1$. Note that this error can
be arbitrarily small by lowering $\zeta_F$ indefinitely, i.e., a very strong measurement.
If phase-squeezed light is injected during the verification stage, we would have
\begin{equation}
U^{\rm add} |_{\rm BAE}\sim (e^{-2q} +2\zeta_x^2)^{1/2} \zeta_F=\sqrt{\frac{\Omega_F^2}{\Omega_q^2e^{2q}}+\frac{2\Omega_F^2}{\Omega_x^2}}.
\end{equation}
Increasing the squeezing factor always improves our verification
sensitivity, with a limit of
\begin{equation}\label{baeU}
U^{\rm add}_{\rm lim}|_{\rm BAE} \sim \Omega_F/\Omega_x= \zeta_x\,\zeta_F,
\end{equation}
which can be much lower than unity in the case of future GW detectors or any
low-noise measurement device.

Had we not evaded the back-action noise, we would have $\sqrt{N_F}$ in the
place of $\zeta_F$, which means $\delta x_V \delta p_V$ would be Heisenberg-limited ---
unless different squeezing factors are assumed. For low squeezing (i.e., $e^{\pm
2q}$ larger than both $\zeta_x$ and $\zeta_F$), we need phase
squeezing for $\hat x$ observation, amplitude squeezing for $\hat
p$ observation, with
\begin{equation}
U^{\rm add}|_{\rm without\,BAE} \sim e^{-q}\,,
\end{equation}
which is a significant factor ($1/\zeta_F$) worse than the BAE
scheme. Even though there exists an optimal squeezing factor that
this scheme can apply and yields
\begin{equation}
U^{\rm add}_{\rm opt}|_{\rm without\,BAE} \sim \zeta_x\,,
\end{equation}
yet it is still worse than the limiting situation of the BAE
scheme [cf. Eq. \eqref{baeU}] by a factor of $1/\zeta_F\,(\gg 1)$.

\section{The conditional quantum state and its evolution}\label{III0}
\label{III0}

The previous order-of-magnitude estimates provide us a qualitative picture of
a MQM experiment, especially in the free-mass regime where future GW detectors
are operating. As long as $\zeta_F$ and $\zeta_x$ are smaller than unity, namely, the
classical noise goes below the SQL around the most sensitive frequency band
$(\Omega\sim\Omega_q)$ of the measurement device,
not only can we prepare a nearly pure quantum state, but also can we make a sub-Heisenberg
tomography of the prepared state. In this and following sections, we will provide
more rigorous treatments directly by analyzing the detailed dynamics of the system.

\subsection{The conditional quantum state obtained from Wiener filtering}
The rigorous mathematical treatment of state preparation has been given in Ref. \cite{state_pre}.
The main idea is to treat the conditional quantum state preparation as a classical filtering problem,
which is justified by the fact that the outgoing optical quadratures $\hat b_{1,2}$ at different times
commute with each other, the same as a classical random process. For such a Gaussian
linear system, the Wiener filter, satisfying the minimum mean-square error criterion, allows us to
obtain an optimal estimate for the quantum state of the oscillator, i.e., the conditional quantum state. Based upon the measurement data $y(t)\;(t<0)$, conditional means for oscillator position and momentum at $t=0$ can be constructed as [cf. Eq. (14) of Ref. \cite{state_pre}]
\begin{align}
x^{\rm cond}(0)\equiv \langle \hat x(0)\rangle^{\rm cond}&=\int _{-\infty}^{0}dt\, K_x(-t) y(t),\\
p^{\rm cond}(0)\equiv \langle \hat p(0)\rangle^{\rm cond}&=\int _{-\infty}^{0}dt\, K_p(-t) y(t).
\end{align}
Here $K_{x}$ and $K_{p}$ are causal Wiener filters. The covariance matrix is given by [cf. Eq. (15) of Ref. \cite{state_pre}]
\be
{\bf V}^{\rm cond}_{o_io_j}(0)= \langle\hat o_i(0)\hat o_j(0) \rangle^{\rm cond}_{\rm sym}-\langle\hat o_i(0)\rangle^{\rm cond}\langle\hat o_j(0)\rangle^{\rm cond},
\ee
where $i,j=1,2$ and $\hat o_1, \hat o_2$ denote $\hat x, \hat p$, respectively.
In the free-mass regime, we showed that [cf. Eq.~(52)--(54) in Ref.~\cite{state_pre}]:
\begin{equation}
\mathbf{V}^{\rm cond}(0) = \left[
\begin{array}{cc}
N_F^{\frac{1}{4}}N_x^{\frac{3}{4}} \sqrt{2}\delta x_q^2 &
N_F^{\frac{1}{2}}N_x^{\frac{1}{2}}\hbar/2 \\
N_F^{\frac{1}{2}}N_x^{\frac{1}{2}}\hbar/2 &
N_F^{\frac{3}{4}}N_x^{\frac{1}{4}} \sqrt{2}\delta p_q^2
\end{array}
\right]. \label{Vc}
\end{equation}
With conditional means and variances, the Wigner function or equivalently the conditional
quantum state is uniquely defined [cf. Eq. \eqref{wigf}].
Correspondingly, purity of the conditional quantum state is quantified by
\be
U(0)=\frac{2}{\hbar}\sqrt{\det {\bf V}^{\rm cond}(0)}=N_xN_F.
\ee
This simply justifies the order-of-magnitude result presented in Eq. \eqref{pesti}.

\subsection{Evolution of the conditional quantum state}
In the following discussions, we will analyze how such a conditional quantum state evolves during the evolution stage. On the one hand, this confirms the qualitative results presented in the
subsection \ref{subsec:evo}. On the other hand, it provides a quantitative understanding of the timescale for the later verification stage.

The equations of motion for the oscillator during the evolution stage are given by Eqs. \eqref{11} and \eqref{12} except that there is no radiation pressure, for the light is turned off \footnote{Were the light turned on, the back action can 
still be evaded as long as one measures the amplitude quadrature $\hat a_1$ during this period and take them into account 
during data processing. Since no information of the oscillator position (contained in the phase quadrature of outgoing light) is collected, this is equivalent to the case with light turned off.}.
For simplicity and also a consideration of the case in a realistic experiment, we will assume an oscillator with a high quality factor, i.e., $\omega_m\gg\gamma_m$.
Within a timescale much shorter than $1/\gamma_m$, the oscillator can be well-approximated
as a free oscillator. Correspondingly, the analytical solution to oscillator position
reads
\begin{equation} \label{xq}
\hat x (t) = \hat x_q (t) + \int_0^\infty {\rm d} t'\, G_x (t -
t')\hat \xi_F (t')\,.
\end{equation}
Here the free quantum oscillation $\hat x_q(t)$ of the oscillator is given by  Eq. \eqref{position}.
We have defined the Green's function as
\begin{equation}
G_x (t) = \Theta(t)\, \frac{\sin (\omega_m\,
t)}{m\,\omega_m}\,,
\end{equation}
with $\Theta(t)$ denoting the Heaviside function.

Given an evolution duration of $\tau_E$, from Eq. \eqref{Fthc} and Eq. \eqref{xq}
the corresponding covariance matrix evolves as
\begin{align} \label{Vtau}{\bf
V}(\tau_E) &= {\bf R}_{\Phi}^{T}\ {\bf V}^{\rm cond}(0)\ {\bf R}_{\Phi}
\nonumber \\ & + \frac{S_F^{\rm
th}}{8\,m^2\omega_m^3}\left[\begin{array}{cc}2\,\Phi-\sin
2\,\Phi & 2\,m\,\omega_m\sin^2\Phi \\
2\,m\omega_m\sin^2\,\Phi & m^2\omega_m^2\left(2\,\Phi+\sin
2\,\Phi\right)\end{array}\right]\,,
\end{align}
where $\Phi\equiv\omega_m\,\tau_E$ and the rotation matrix ${\bf R}_{\Phi}$ is given by
\be
{\bf R}_{\Phi}=\left[\begin{array}{cc}\cos\Phi&-m\,\omega_m
\sin\Phi
\\ ({m\,\omega_m})^{-1}{\sin\Phi}&\cos{\Phi}\end{array}\right]\,.
\ee
The first term in Eq. \eqref{Vtau} represents a rotation of the covariance matrix
${\bf V}^{\rm cond}(0)$ due to the free quantum oscillation of the oscillator; the second
term is contributed by thermal decoherence
which causes an increase in the uncertainty.

In the free-mass regime and the case of $\omega_m\tau_E\ll 1$,
elements of the covariance matrix can be expanded as series of
$\Phi$. Up to the leading order in $\Phi$, we obtain
\begin{align}
V_{xx} (\tau_E) &= V_{xx}^{\rm cond} + \frac{4\delta x_q^2}{\hbar}
V_{xp}^{\rm cond}\Omega_q\tau_E +\frac{\delta x_q^2}{\delta p_q^2} V_{pp}^{\rm cond}(\Omega_q\tau_E)^2\nonumber \\
&+ 2\delta x_q^2\zeta _F^2\,
\frac{(\Omega_q\tau_E)^3}{3}\,, \label{Eq:Vxxt} \\
V_{xp} (\tau_E) &= V_{xp}^{\rm cond} + \frac{\hbar}{2\delta p_q^2} V_{pp}^{\rm
cond}\Omega_q\tau_E
+ \frac{\hbar}{2}\, \zeta _F^2(\Omega_q\tau_E)^2\,, \label{Eq:Vxpt}\\
V_{pp} (\tau_E) &= V_{pp}^{\rm cond} + 2\delta p_q^2 \zeta
_F^2\Omega_q\tau_E\, \label{Eq:Vppt}
\end{align}
with $V^{\rm cond}_{xx,xp,pp}$ denoting the elements of ${\bf V}^{\rm cond}(0)$. Up to the leading order in $\Omega_q\tau_E$, the uncertainty product of the resulting quantum state is
\begin{align}\label{utaue}
U(\tau_E)=\frac{2}{\hbar}\sqrt{\det {\bf V(\tau_E)}}\approx
U(0)+\frac{V_{xx}^{\rm cond}}{\delta x_q^2}(\tau_E/\tau_F)^2
\end{align}
with $\tau_F$ defined in Eq. \eqref{ts}.
The second term is contributed by the thermal decoherence and can be viewed as $U^{\rm th}(\tau_E)$.
Those formulas recover the results in Eqs. \eqref{29} -- \eqref{33} but with precise numerical factors.
As we can conclude from Eq. \eqref{utaue}, in order for a sub-Heisenberg tomography to be possible, the later verification stage
should finish within a timescale of $\tau_F$ after which the contribution from the thermal noise gives $U^{\rm th}(\tau_F)\sim 1$.

\section{State verification in the presence of Markovian Noises}\label{III}
\label{markovian} In this section, we will treat the followup state
verification stage with Markovian noises in details. This can justify the
order-of-magnitude estimate we have done in the subsection \ref{subsec:verify}.
In addition, we will show explicitly how to construct the optimal
verification scheme that gives a sub-Heisenberg accuracy.

\subsection{A time-dependent homodyne detection and back-action-evading (BAE)}

In this subsection, we will analyze the time-dependent homodyne detection which
enables us to probe mechanical quadratures. We will further show how the BAE scheme can be constructed.
The BAE scheme is optimal only when there is no readout loss ($\eta=0$). We will consider
more general situations and derive the corresponding optimal verification scheme in the
next subsection.

The equations of motion for the oscillator
during the verification stage ($t>\tau_E$) are given by Eqs. \eqref{11} and \eqref{12}.
The corresponding solution to oscillator position is different from Eq. \eqref{xq} due to the
presence of the back-action noise which starts to act on the oscillator at $t=\tau_E$.
Specifically, it reads
\begin{equation} \label{xqt}
\hat x (t) = \hat x_q (t) + \int_{\tau_E}^\infty {\rm d} t'\, G_x (t -
t')[\alpha\,\hat a_1(t')+\hat \xi_F (t')]\,.
\end{equation}
Here the free quantum oscillation $x_q(t)$ is the signal that we seek to probe during the verification stage. For optical quadratures, the equations of motion are given by Eqs. \eqref{13} and \eqref{14}.
From those equations, we notice that among the
outgoing fields: $\hat b_1$ is pure noise, while $\hat b_2$ contains both signal
$\hat x_q(t)$ and noise.  In order to highlight this, we rewrite $\hat b_{1,2}$ as
\begin{eqnarray}
\label{eq:b1}
\hat b_1(t) &=& \sqrt{1-\eta}\,\hat n_1(t)+\sqrt{\eta}\,\hat a_1(t)\equiv \delta \hat b_1(t) \,,\\
\hat b_2(t) &=& \delta \hat b_2(t) + \sqrt{1-\eta}\,({\alpha}/{\hbar})\, \hat
x_q(t)
\end{eqnarray}
with [cf. Eq. \eqref{xqt}]
\begin{align} \label{eq:db2}
\delta \hat b_2 (t) &\equiv \sqrt{\eta}\,\hat n_2(t)+\sqrt{1-\eta}\Big\{\hat a_2(t)
+\frac{\alpha}{\hbar}\,\hat \xi_{x}(t) \nonumber \\ & \quad +
\frac{\alpha}{h}\int_{\tau_E}^{\infty}{\rm d} t'\,
G_x(t-t')\,[\alpha\, \hat a_1(t') + \hat\xi_{F}(t')]\Big\}\,.
\end{align}
In this way, we can directly see that $\hat a_1$ which causes the back action is contained
in both the amplitude quadrature $\hat b_1$ and the phase quadrature $\hat b_2$. Therefore,
by measuring an appropriate combination of the two output quadratures, we will be able to
remove effects of the back-action noise that is imposed onto the oscillator during the
verification process at $t>\tau_E$. Searching for
such an optimal combination is the main issue to be addressed in this section.

\begin{figure}
\includegraphics[width=0.45\textwidth, bb=0 0 193 77,clip]{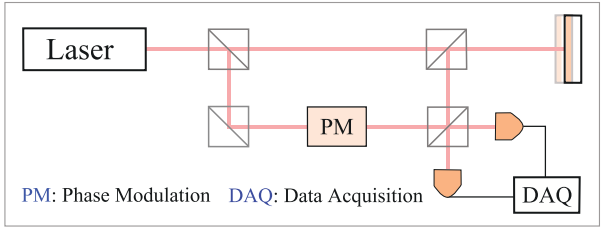}
\caption{ (Color online) A schematic plot of time-dependent homodyne detection. The phase modulation of
the local oscillator light varies in time. \label{config}}
\end{figure}

As mentioned in the introduction part, to probe mechanical quadratures and their distributions,
a time-dependent homodyne detection needs to be applied [cf. Eq. \eqref{intest}]. Specifically,
the outgoing optical field
\be \hat B_{\rm out}(t)=\hat
b_1(t)\cos\omega_0 t+\hat b_2(t)\sin\omega_0 t
\ee
at $t>\tau_E$ is mixed with a strong local-oscillator light $L(t)$ whose
phase angle $\phi_{os}$ is time-dependent as shown schematically in Fig. \ref{config}, namely,
\begin{equation}
L(t) = L_0\cos[\omega_0\,t-\phi_{os}(t)]
\end{equation}
with $L_0$ a time-independent constant. Through a
low-pass filtering (with a bandwidth much smaller than $\omega_0$) of the beating signal, the resulting
photocurrent is
\begin{align}\nonumber
\hat i(t) &\propto 2\overline{\hat B_{\rm out}(t) L(t)} \\&=
L_0\,\hat b_1(t)\cos\phi_{os}(t)+L_0\,\hat b_2(t)\sin\phi_{os}(t)\,,
\end{align}
where the overline means averaging over many optical-oscillation periods. Note that Heisenberg
operators for the photocurrent at different time commute with each other, i.e.,
\begin{equation}\label{43}
[\hat i(t), \hat i(t')]=0\,,
\end{equation}
and are therefore {\it simultaneously measurable}, as obviously expected.  Based on the measurement
results of $\hat i(t)$ from $\tau_E$ to $T_{\rm int}$, we can construct the following weighted
quantity $\hat Y$ with a weight function $W(t)$:
\begin{equation}
\label{eq:Y} \hat Y=\int_{0}^{T_{\rm int}} \Theta(t-\tau_E) W(t) \hat i(t) dt \equiv
( g_1|\hat b_1)+( g_2|\hat b_2)\,.
\end{equation}
Here the Heaviside function $\Theta(t-\tau_E)$ manifests the fact that
the verification stage starts at $t=\tau_E$ and we have introduced the scalar product of two vectors $|A)$ and $|B)$ in the ${\cal L}^2[0, T_{\rm int}]$ space as the following: 
\begin{equation}
( A|B) \equiv \int_{0}^{T_{\rm int}} A(t) B(t) dt\,.
\end{equation}
Besides, we have defined filtering functions $g_1$ and $g_2$ as
\begin{align}
g_1(t)&\equiv \Theta(t-\tau_E)W(t)\cos\phi_{os}(t),\\
g_2(t)&\equiv \Theta(t-\tau_E)W(t)\sin\phi_{os}(t).
\end{align}
Since all the data can in principle be digitalized and
stored in hardwares, the weight function $W(t)$ can be realized digitally during data processing. In
addition, an overall re-scaling of $g_{1,2}(t) \rightarrow C_0\,g_{1,2}(t)$ with $C_0$ a time-independent
constant does not affect the verification performance, and that there are multiple ways of achieving
a particular set of $g_{1,2}(t)$, by adjusting the phase $\phi_{os}(t)$ of the local oscillator
and the weight function $W(t)$.

In light of Eqs. \eqref{eq:b1} -- \eqref{eq:db2}, we decompose the weighted quantity  $\hat Y$ [cf. Eq. \eqref{eq:Y}] as a signal $\hat Y_s$ and a noise part $\delta \hat Y$, namely,
\begin{equation}
\hat Y =  \hat Y_s +\delta \hat Y.
\end{equation}
They are given by
\begin{eqnarray}
\hat Y_s &=&  \sqrt{1-\eta}\,({\alpha}/{\hbar})\,( g_2|\hat x_q), \nonumber \\
\delta \hat Y &=& ( g_1|\delta \hat b_1) +( g_2|\delta \hat b_2).
\end{eqnarray}
Since an overall normalization of $g_{1,2}$ will not affect the signal-to-noise ratio as mentioned, we can impose, mathematically, that
\begin{equation}
\label{eqg} ( g_2 |f_1)  = \cos\zeta\,,\quad ( g_2 | f_2) =
\sin\zeta
\end{equation}
with
\be
f_1(t)\equiv \cos\omega_m t, \quad f_2(t)\equiv (\Omega_q/\omega_m)\sin\omega_m t
\ee in the coordinate representation. The signal part can then be rewritten as
\be\label{ys}
\hat Y_s=\sqrt{1-\eta}\,({\alpha}/{\hbar})\delta x_q\left[\hat x_0\cos\zeta+\hat p_0\sin\zeta\right]\,,
\ee
where we have introduced normalized oscillator position and momentum as $\hat x_0\equiv \hat x(\tau_E)/\delta x_q$ and $\hat p_0\equiv \hat p(\tau_E)/\delta p_q$.
In such a way, a mechanical quadrature of $\hat X_\zeta$ will be probed [cf. Eq. \eqref{ROquad}].
For the noise part, more explicitly, we have [cf. Eqs. \eqref{eq:b1}--\eqref{eq:db2}]
\begin{align}
\label{dYdecomp}
\delta \hat Y &=( g_1|\sqrt{\eta}\,\hat n_1+\sqrt{1-\eta}\,\hat a_1)+ ( g_2|\sqrt{\eta}\,\hat n_2 +\sqrt{1-\eta}\, \hat a_2)
\nonumber \\
&+ \sqrt{1-\eta}\, ({\alpha^2}/{h}) ( g_2 |\mathbf{G}_x | \hat a_1) \nonumber \\
&+ \sqrt{1-\eta}\,  ({\alpha}/{\hbar})[ ( g_2|\mathbf{G}_x|\hat \xi_{F}) +( g_2|\hat \xi_x) ]\,,
\end{align}
where the integration with $G_x(t-t')$ has been augmented into applying a linear operator
$\mathbf{G}_x$ in the ${\cal L}^2[0,\,T_{\rm int}]$ space. In the above equation, terms on
the first line is the shot noise, and the term on the second line is the back-action noise,
while terms on the third line are the classical-force and sensing noises.

The optimal $g_1(t)$ and $g_2(t)$ that give a sub-Heisenberg accuracy for each quadrature
will be rigorously derived for general situations in the next section. If $\hat a_{1}$ and
$\hat a_2$ are uncorrelated and there is no readout loss with $\eta=0$, an optimal choice for $g_1$ would be obvious to cancel the entire contribution from the
back-action noise term (proportional to $\hat a_1$). This is equivalent to impose, mathematically, that
\begin{equation}
(g_1|\hat a_1)  +({\alpha^2}/{h}) ( g_2 |\mathbf{G}_x | \hat a_1)
=0
\end{equation}
or
\begin{equation}\label{bae47}
|g_1) + ({\alpha^2}/{h})\,\mathbf{G}_x^{\rm adj} |g_2) =0\,,
\end{equation}
where $\mathbf{G}_x^{\rm adj}$ is the adjoint of $\mathbf{G}_x$. Physically, this
corresponds to bringing in a piece of shot noise $(g_1|\hat a_1) $ to cancel the
back-action noise $(\alpha^2/h) ( g_2 |\mathbf{G}_x |\hat a_1) $ --- therefore achieving
a only shot-noise-limited measurement. In the coordinate representation, Eq. \eqref{bae47} can be written out more explicitly as
\be\label{bae}
g_1(t)+(\alpha^2/\hbar)\int_t^{T_{\rm int}}dt' G_x(t'-t)
g_2(t')=0, \ee
which agrees exactly with the variational-type BAE measurement scheme first
investigated by Vyatchanin {\it et al.} \cite{bae}. It is suitable for detecting
signals with  {\it known arrival time}. For stationary signals, one would prefer
frequency-domain variational techniques proposed by Kimble {\it et al.}~\cite{klmtv},
which evades the back-action noise for all possible signals as long as they are
Gaussian and stationary.

As realized by Kimble {\it et al.}~\cite{klmtv} in their frequency-domain treatment,
when  the readout loss is significant (large $\eta$) and when the back-action noise is
strong (large $\alpha$), the variational approach becomes less effective, because in
such a case, the magnitude of $g_1$ required to bring enough $\hat a_1$ to cancel
the back-action noise would also introduce significant noise $\hat n_1$ [cf. Eq. \eqref{dYdecomp}]. This reasoning apparently leads to a trade-off between
the need to evade back action and the need to minimize loss-induced shot noise
--- such an optimization will be made in the next section.

\subsection{Optimal verification scheme and covariance matrix for the added noise: formal derivation}\label{IIIC}

Imposing the BAE condition [cf. Eq. \eqref{bae}] does not specify the shape of $g_2$, nor does
Eq.~\eqref{eqg}, and we have further freedom in choosing $g_2$ that minimizes the noise in
measuring a particular quadrature of $\hat X_{\zeta}$. In addition, in the presence of
readout loss with $\eta\neq0$, totally evading back action is not the obvious optimum as
mentioned. Therefore we need to optimize $g_{1}$ and $g_2$ simultaneously. In this section,
we first carry out this procedure formally, and apply to the Markovian-noise budget in the
next subsection.

The total $\hat x_q$-referred noise in the weighted estimator $\hat Y$ can be written as [cf. Eqs. \eqref{ys} and \eqref{dYdecomp}]
\begin{align}\nonumber
\sigma^2[g_{1,2}]&=\frac{\hbar^2}{(1-\eta)\alpha^2\delta x_q^2}\langle\delta\hat Y \delta\hat Y\rangle_{\rm sym}\\&=\frac{2}{(1-\eta)\Omega_q}\sum_{i,j=1}^2( g_i|\mathbf{C}_{ij}|g_j) ,
\end{align}
where correlation functions ${\bf C}_{ij}$ among the noises are the following:
\begin{equation}\label{66}
{\bf C}_{ij}(t,t') \equiv \langle \delta \hat b_i(t) \delta \hat
b_j(t')\rangle_{\rm sym} \,,\quad (i,j=1,2)\,.
\end{equation}
The optimal $g_{1,2}(t)$ that minimize $\sigma^2$ can be obtained
through the standard constraint variational method. For this, we define an
effective functional as
\ba
{\cal J}_{\rm eff}&=&(1-\eta)({\Omega_q}/{4})\sigma^2[g_{1,2}] -\mu_1 ( f_1|g_2)-\mu_2 ( f_2|g_2) \nonumber \\
&=&\frac{1}{2}\sum_{i,j}(g_i|\mathbf{C}_{ij}|g_j)  - ( \mu_1 f_1+
\mu_2 f_2 |g_2),
\ea
where $\mu_1$ and $\mu_2$ are Lagrange multipliers due to the normalization constraints
in Eq. \eqref{eqg}. Requiring the functional derivative of ${\cal J}_{\rm eff}$ with respect
to $g_1$ and $g_2$ equal to zero, we obtain
\ba \label{52}
\mathbf{C}_{11} |g_1) + \mathbf{C}_{12} |g_2)  &=&0\,,\\\label{53}
\mathbf{C}_{21} |g_1)+ \mathbf{C}_{22} |g_2)  &=&| \mu_1 f_1 +
\mu_2 f_2)\,.
\ea
Here ${\bf{C}}_{ij}$ should be viewed as operators in the ${\cal L}^2[0,\,T_{\rm int}]$ space.
This leads to formal solutions to $g_{1,2}$, namely,
\begin{eqnarray}\label{47}
|g_1) &=& -\mathbf{C}_{11}^{-1}
\mathbf{C}_{12}|g_2)\,,\\\label{48}
 |g_2) &=&
\mathbf{M}| \mu_1 f_1 + \mu_2 f_2)\,,
\end{eqnarray}
where we have defined
\begin{equation}\label{M}
\mathbf{M} \equiv
\left[\mathbf{C}_{22}-\mathbf{C}_{21}\mathbf{C}_{11}^{-1}
\mathbf{C}_{12}\right]^{-1}\,.
\end{equation}
Re-imposing Eqs.~\eqref{eqg}, those unknown Lagrange multipliers $\mu_{1,2}$
can be solved, which are related to $\zeta$ by
\begin{equation}
\label{50} \left[
\begin{array}{cc}
( f_1 |\mathbf{M}|f_1) & ( f_1 |\mathbf{M} |f_2)\\
( f_2 |\mathbf{M} |f_1) & ( f_2 |\mathbf{M} |f_2)
\end{array}
\right] \left[
\begin{array}{c}
\mu_1 \\
\mu_2
\end{array}
\right] = \left[
\begin{array}{c}
\cos\zeta \\
\sin\zeta
\end{array}
\right].
\end{equation}
Correspondingly, the minimum $\sigma^2_{\rm min}$ has the following quadratic form:
\begin{eqnarray}
\sigma^2_{\rm min}
=
[\cos\zeta\;\sin\zeta]\mathbf{V}^{\rm add}_{\rm norm} \left[
\begin{array}{c}
\cos\zeta \\
\sin\zeta
\end{array}
\right].
\end{eqnarray}
Here normalized $\mathbf{V}^{\rm add}_{\rm norm}$ is a $2\times2$ covariance matrix,
and it is given by
\begin{equation}
\label{Vadd} \mathbf{V}^{\rm add}_{\rm norm}= \frac{2}{(1-\eta)\Omega_q}\left[
\begin{array}{cc}
( f_1 |\mathbf{M}|f_1) & ( f_1 |\mathbf{M} |f_2)\\
( f_2 |\mathbf{M} |f_1) & ( f_2 |\mathbf{M} |f_2)
\end{array}
\right]^{-1}.
\end{equation}
It relates to the initial definition of the covariance matrix for the added verification noise
[cf. Eq. \eqref{vadef}] simply by
\be
{\bf V}^{\rm add}={\rm Diag}[\delta x_q,\,\delta p_q] {\bf V}^{\rm add}_{\rm norm}{\rm Diag}[\delta x_q,\, \delta p_q].
\ee

Due to the linearity in Eqs.~\eqref{53} and \eqref{50},  the optimal $g_{1,2}$ for a given
quadrature $\zeta$ can also be rewritten formally as
\begin{equation}\label{63}
g_{1,2}^{\zeta} = g_{1,2}^X \cos\zeta+ g_{1,2}^P\sin\zeta\,,
\end{equation}
with $g_{1,2}^{X}\equiv g^{\zeta}_{1,2}(0)$ and $g_{1,2}^{P}\equiv g^{\zeta}_{1,2}(\pi/2)$.
Such $\zeta$-dependence of $g_{1,2}$ manifests the fact that a sub-Heisenberg tomography requires
different filtering functions, or equivalently different measurement setups, for different quadratures.

\subsection{Optimal verification scheme with Markovian noise}\label{IIID}
Given Makovian noises, the corresponding correlation functions for the output
noise $\delta\hat b_i$ can be written out explicitly as [cf. Eqs. \eqref{Fthc},  \eqref{FBAc}, \eqref{Sxthc}, and \eqref{66}]
\begin{align}
C_{11}(t,t')&=\frac{\eta+(1-\eta)e^{2q}}{2}\delta(t-t'),\\\label{32}
C_{12}(t,t')&=C_{21}(t',t)=(1-\eta)\frac{e^{2q}\alpha^2}{2\hbar}G_x(t'-t),\\\nonumber
C_{22}(t,t')&=\frac{\Lambda^2}{4}\delta(t-t')+(1-\eta)\frac{\alpha^4}{\hbar^2}
\left(\frac{e^{2q}}{2}+\zeta_F^2\right)\\&
~~~~ {\int_0^{\infty}}dt_1
G_x(t-t_1)G_x(t'-t_1),
\end{align}
with  $\Lambda\equiv\sqrt{2[\eta+(1-\eta)(e^{-2q}+2\zeta_x^2)]}$.
Plugging these $C_{ij}$ into Eq. \eqref{47} and \eqref{48}, we can
obtain the equations for the optimal filtering functions $g_1$ and
$g_2$. Specifically, for $g_1$, we have [cf. Eq. \eqref{47}]
\be \label{59}
g_1(t)+\frac{(1-\eta)e^{2q}}{\eta+(1-\eta)e^{2q}}\frac{\alpha^2}{\hbar}\int_t^{T_{\rm
int}}dt' G_x(t'-t) g_2(t')=0.
\ee
For $g_2$, by writing out $\bf M$ explicitly, it gives [cf. Eq. \eqref{48}]
\begin{align}\nonumber
\frac{\Lambda^2}{4}g_2(t)&+\zeta_F'^{\,2}\frac{\alpha^4}{\hbar^2}
\iint_0^{T_{\rm
int}}dt'dt_1G_x(t-t_1)G_x(t'-t_1)g_2(t')\\&=\mu_1 f_1(t)+\mu_2 f_2(t)\,, \label{60}
\end{align}
where we have introduce $\zeta_F'$, which is given by
\be\label{zF}
\zeta_F'\equiv\left[\frac{\eta(1-\eta) e^{2q}}{2[\eta+(1-\eta)e^{2q}]}+(1-\eta)\zeta_F^2\right]^{1/2}\approx
\left[\frac{\eta}{2}+\zeta_F^2\right]^{1/2}
\ee
and it is equal to $\zeta_F$ for no readout loss.
Although here $g_1$ is still defined from $g_2$, the optimal
verification strategy does not totally evade the back action, as is
manifested in the term proportional to $\eta$ inside the bracket
of Eq. \eqref{zF}. In the limit of no readout loss with $\eta=0$, it
is identical to the BAE condition in Eq. \eqref{bae} .
Typically, we have $1\%$ readout loss $\eta=0.01$, squeezing
$e^{2q}=10$ and $\zeta_F=0.2$, this readout loss will only shift
$\zeta_F$ by $6\%$, which is negligible. However, if the thermal noise
further decreases and/or the measurement strength increases, the effect of readout
loss will become significant, entering in
a similar way as the frequency-domain variational measurement
proposed by Kimble {\it et al.} \cite{klmtv}.

\begin{figure}
\includegraphics[width=0.47\textwidth, bb=0 0 344 182,clip]{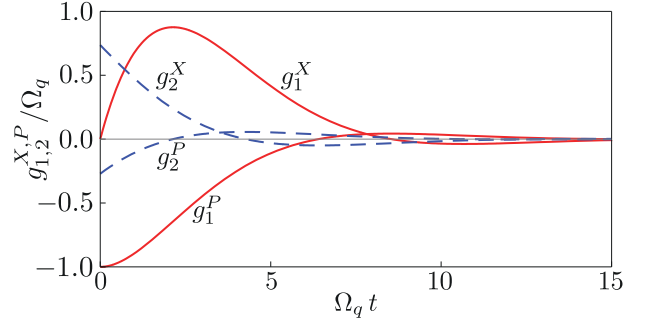}
\caption{(Color online) Optimal filtering functions $g_1$ (solid curve) and $g_2$
(dashed curve) in the presence of Markovian noises. We have assumed
$\Omega_q/2\pi=100$ Hz, $\zeta_x=\zeta_F=0.2$, $\eta=0.01$ and
vacuum input ($q=0$). For clarity, the origin of the time axis has been
shifted from $\tau_E$ to 0.} \label{gxgp}
\end{figure}
The above integral equations for optimal $g_1$ and $g_2$ can be solved
analytically as elaborated in the Appendix \ref{AppB}, which in turn gives
$\bf M$ and the corresponding ${\bf V}^{\rm add}$ [cf. Eqs. \eqref{M} and \eqref{Vadd}].
In the free-mass regime with $\Omega_q\gg\omega_m$, closed forms for optimal
$g_1$ and $g_2$ can be obtained, which, in terms of $g_{1,2}^{X,P}$ [cf. Eq. \eqref{63}],
are given by
\begin{align}
g_1^X&=g_1|_{\zeta=0}=(\Omega_q/\chi)\,e^{-\Omega_q\chi\,t}\sin\Omega_q\chi\, t;\\
g_1^{P}&=g_1|_{\zeta=\frac{\pi}{2}}=-\sqrt{2}\,\Omega_q\,e^{-\Omega_q\chi
\,t}\sin\left(\Omega_q\chi \,t+\frac{\pi}{4}\right),
\end{align}
and
\begin{align}
g_2^X&=g_2|_{\zeta=0}=2\,\Omega_q\chi\,e^{-\Omega_q\chi\, t}\cos\Omega_q\chi\, t;\\
g_2^{P}&=g_2|_{\zeta=\frac{\pi}{2}}=2\sqrt{2}\,\Omega_q\chi^2\,e^{-\Omega_q\chi
\,t}\sin\left(\Omega_q\chi t-\frac{\pi}{4}\right),
\end{align}
with
$\chi\equiv[\zeta_F'^{\,2}/\Lambda]^{1/2}$. The corresponding verification timescale is set by $\tau_V=(\chi\,\Omega_q)^{-1}$ and $\tau_q<\tau_V<\tau_F$.
To illustrate the behavior of the optimal filtering functions, we
show $g_{1,2}^{X,P}$ in Fig. \ref{gxgp}. As we can see, the verification
process finishes after several $\tau_q$, i.e., in a timescale of $\tau_V$.

\begin{figure}
\includegraphics[width=0.25\textwidth, bb=0 0 300 300,clip]{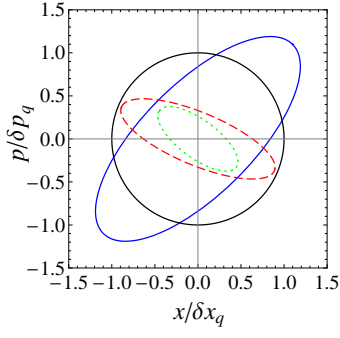}
\caption{(Color online) The uncertainty ellipse for the added verification noise
in the presence of Markovian noises. We assume $\zeta_x=\zeta_F=0.2$, vacuum input
(Dashed curve), $\zeta_x=\zeta_F=0.2$ and 10\,dB squeezing (Dotted
curve). For contrast, we also show the Heisenberg limit in a unit circle and the
ideal conditional quantum state in solid ellipse. \label{contour}}
\end{figure}

The corresponding covariance matrix ${\bf V}^{\rm add}$ for the added verification noise is given
by
\begin{equation}\label{covmat}
\mathbf{V}^{\rm add}=\frac{1}{1-\eta}\left[
\begin{array}{cc}
\Lambda^{\frac{3}{2}}\zeta_F'^{\,\frac{1}{2}}\delta x_q^2  &  -\Lambda\zeta_F'\hbar/2\\
-\Lambda\zeta_F'\hbar/2 &
2\Lambda^{\frac{1}{2}}\zeta_F'^{\,\frac{3}{2}}\delta p_q^2
\end{array}
\right]\,.
\end{equation}
A more summarizing measure of the verification accuracy is the uncertainty product of
the added noise ellipse with respect to the Heisenberg limit, namely,
\begin{equation}
U^{\rm add} = \frac{2}{\hbar}\sqrt{\det\mathbf{V}^{\rm add}}
=\frac{\Lambda\,\zeta_F'}{1-\eta}\,.
\end{equation}
In the ideal case with $\eta=0$, this simply recovers the order-of-magnitude estimate given in
the subsection \ref{subsec:verify}. In Fig.~\ref{contour}, we show the uncertainty
ellipse for the added noise in the case of $\zeta_x =\zeta_F=0.2$, readout loss $\eta=1\%$
and with (Green dotted curve) or without (red long-dashed curve) 10\,dB input squeezing.
In comparison, we also plot the Heisenberg limit (unit
circle) and the conditional state obtained through an ideally noiseless state
preparation (blue solid ellipse). As  figure shows, the least
challenging scenario already begins to characterize the
conditional quantum state down to the Heisenberg Uncertainty. In this two
cases, we have $\Lambda=1.48$ and $0.62$ respectively, leading to
\begin{equation}
U^{\rm add} = 0.30 \;\mbox{(vacuum)}\,,\;\;0.12\;(10\,\mbox{dB
squeezing}).
\end{equation}

\section{Verification of Macroscopic Quantum Entanglement}
\label{ent} 
\begin{figure}
\includegraphics[height=0.45\textwidth, bb = 0 0 397 355, clip]{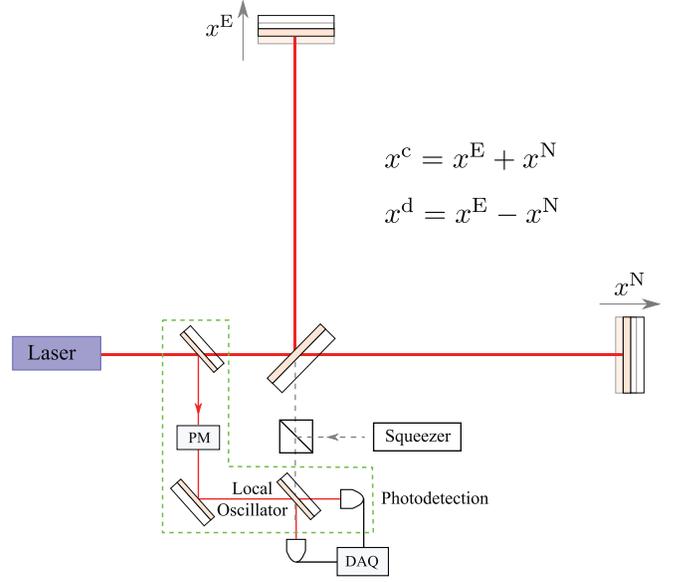}
\caption{(Color online) A schematic plot of advanced interferometric GW detectors
for macroscopic entanglement between test
masses as a test for gravity decoherence.
For simplicity, we have not shown the setup at the bright port, which
is identical to the dark port. } \label{Grav}
\end{figure}

In this section, we will apply our
protocol to verify macroscopic entanglement between test masses in future GW detectors,
which was proposed in Refs. \cite{MRSDC2007, state_pre}. In the
experiment as shown schematically in Fig. \ref{Grav}, measurements
at the bright and dark port of the interferometer continuously collapse the quantum
state of the corresponding common and differential modes of the test-mass motion.
This creates two highly squeezed Gaussian state in both modes. Since the
common and differential modes are linear combinations
of the center of mass motion of test masses in the two arms, namely $\hat x^{\rm c}=\hat x^{\rm E}+\hat x^{\rm N}$ and $\hat x^{\rm
d}=\hat x^{\rm E}-\hat x^{\rm N}$, this will naturally generate quantum entanglement between
the two test masses, which is similar to creating entanglement by mixing two optical squeezed states at the beam splitter
\cite{Furusawa,Bowen}.
\subsection{Entanglement survival time}

To quantify the entanglement strength, we follow Refs. \cite{MRSDC2007, state_pre} by evaluating
the entanglement monotone --- logarithmic negativity defined in Refs. \cite{Vidal, Adesso}. It
can be derived from the covariance matrix for Gaussian-continuous-variable system considered here.
The bipartite covariances among
$(\hat x^{\rm E}, \hat p^{\rm E}, \hat x^{\rm N}, \hat p^{\rm N})$
form the following covariance matrix:
\be
{\mathbf V}=\left[\begin{array}{cc}{\mathbf V}_{\rm EE}&{\mathbf V}_{\rm EN}\\
{\mathbf V}_{\rm NE}&{\mathbf V}_{\rm NN} \end{array}\right], \ee
where
\begin{align}
{\mathbf V}_{\rm EE} &={\mathbf V}_{\rm NN} =\left[\begin{array}{cc}
(V_{xx}^{\rm c}+V_{xx}^{\rm d})/4&(V_{xp}^{\rm c}+V_{xp}^{\rm d})/2
\\(V_{xp}^{\rm c}+V_{xp}^{\rm d})/2&(V_{pp}^{\rm c}+V_{pp}^{\rm d})\end{array}\right],\\
{\mathbf V}_{\rm NE} &={\mathbf V}_{\rm EN}
=\left[\begin{array}{cc}(V_{xx}^{\rm c}-V_{xx}^{\rm
d})/4&(V_{xp}^{\rm c}-V_{xp}^{\rm d})/2\\(V_{xp}^{\rm
c}-V_{xp}^{\rm d})/2&(V_{pp}^{\rm c}-V_{pp}^{\rm
d})\end{array}\right].
\end{align}
The logarithmic negativity $E_{\cal N}$ can then be written as \be
E_{\cal N}=\max[0,-\log_2 2\sigma_-/\hbar], \ee where
$\sigma_-\equiv \sqrt{(\Sigma-\sqrt{\Sigma^2-4\det {\mathbf
V}})/2}$ and $\Sigma\equiv \det {\mathbf V}_{\rm NN}+\det {\mathbf
V}_{\rm EE}-2\det {\mathbf V}_{\rm NE}$. In contrast to Refs.
\cite{MRSDC2007, state_pre}, now the covariance matrix $\bf V$ corresponds to the total
covariance matrix ${\bf V}^{\rm tot}$ after the entire preparation-evolution-verification process.
For Gaussian quantum states, we have [cf.  Eqs. \eqref{Vplus}, \eqref{Vtau} and \eqref{covmat}]
\be\label{Vtot}
{\mathbf V}^{\rm tot}={\mathbf V}(\tau_E)+{\mathbf V}^{\rm add}.
\ee

\subsection{Entanglement Survival as a Test of Gravity Decoherece}
\begin{figure}
\includegraphics[width=0.45\textwidth, bb=0 0 291 150,clip]{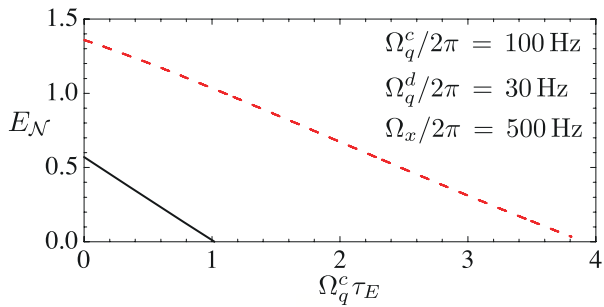}
\caption{(Color online) Logarithmic negative $E_{\cal N}$ as a function of the
evolution duration $\tau_E$, which indicates
how long the entanglement survives. The solid
curve corresponds to the case where $\Omega_F/2\pi=20\,{\rm Hz}$
and the dashed curve for $\Omega_F/2\pi=10\,{\rm Hz}$. To maximize
the entanglement, the common mode is 10 dB phase squeezed at $t>\tau_E$
and $t<0$ while the differential mode is 10 dB amplitude squeezed
at $t<0$ and switching to 10 dB phase squeezed at $t>\tau_E$.}
\label{EN}
\end{figure}

When $\tau_E$ increases, the thermal decoherence will increase the uncertainty [cf. Eqs. \eqref{Vtau} and
\eqref{Vtot}] and eventually the entanglement vanishes, which indicates how long the quantum
entanglement can survive. Survival of such quantum entanglement can help us to understand whether
there is any additional decoherence effect, such as {\it Gravity decoherence} suggested by Di\'{o}si and Penrose \cite{Diosi, Penrose}.
According to their models, quantum superpositions vanish within a timescale  of  $\hbar/E_G$. Here
$E_G$ can be (a) self-energy of the mass-distribution-difference,
namely
\be E_G^{(a)}={\mbox{$\int$}}d{\bf x}d{\bf y}\,{G[\rho({\bf
x})-\rho'({\bf x})][\rho({\bf y})-\rho'({\bf y})]}/{r}
\ee
with $\rho$ denoting the mass density distribution and $r\equiv |{\bf
x}-{\bf y}|$; Alternatively, it can be (b) spread of mutual
gravitational energy among components of the quantum
superposition, namely
\be E_G^{(b)}=\int d{\bf x}d{\bf
y}\,{G\rho({\bf x})\rho'({\bf y})}\,\delta r/r^{3/2}.
\ee
with $\delta r$ denoting the uncertainty in location.
For the prepared test-mass quantum states with width of $\delta
x_q$, we have
\begin{equation}
\tau_G^{(a)} \approx \Omega_q/(G\rho)\,,\; \tau_G^{(b)} \approx
\hbar^{1/2} L^2\Omega_q^{1/2}/(Gm^{3/2})\,.
\end{equation}
where $L$ is the distance between two test masses. Plugging the typical values for
LIGO mirrors with $\rho=2.2\,g/{\rm cm}^3$, the separation between two input test masses,
$L\approx 10\,{\rm m}$
 and $m=10\,{\rm kg}$, we have
\be \tau_G^{a}=4.3\times
10^9\,{\rm s},\quad\tau_G^{b}=1.2\times 10^{-5}\,{\rm s}.
\ee
It is therefore quite implausible to test model (a); while for model
(b), $\Omega_q \tau_G^{(b)}$ is less 0.01 with $\Omega_q/2\pi=
100$ Hz. In Fig. \ref{EN}, we show the entanglement survival as a
function of evolution duration. As we can see, the model (b) of
gravity decoherence can easily be tested, for the entanglement can
survive for several times of the measurement timescale
$\tau_q$, which is much longer than the predicted $\tau_G^{(b)}$.

\section{Conclusions}
\label{con} We have investigated in great details of a followup verification
stage after the state preparation and evolution. We have showed the necessity of a sub-Heisenberg
verification accuracy in probing the prepared conditional quantum state, and how
to achieve it with an optimal time-domain homodyne detection.
Including this essential building block --- a sub-Heisenberg verification,
we are able to outline a complete procedure of a
three-staged experiment for testing macroscopic quantum mechanics.
In particular, we have been focusing on the relevant free-mass regime and have applied
the techniques to discuss MQM experiments with future GW detectors.
However, the system dynamics that have been considered describe general cases with a
high-Q mechanical oscillator coupled to coherent optical fields. To this respect, we note that in our results for Markovian systems only depend on the ratio between various noises and the SQL,
and therefore carries over directly to systems with other scales. In addition, the
Markovian assumption applies more accurately to smaller-scale systems which operate in
higher frequencies.

\acknowledgments We thank Farid Khalili and 
all the members of the AEI-Caltech-MIT
MQM discussion group for very fruitful discussions. We thank K. S.
Thorne for initiating this research project, and V.B. Braginsky
for important critical comments. Research of Y.C.,  S.D., H.M.-E. and K.S. is supported
by the Alexander von Humboldt Foundations Sofja Kovalevskaja Programme,
as well as NSF grants PHY-0653653 and PHY-0601459 and the David and
Barbara Groce startup fund at Caltech. Research of H.R. is supported by the
Deutsche Forschungsgemeinschaft through the SFB No. 407.  H.M. has been
supported by the Australian Research Council and the Department of Education,
Science and Training.  H.M. would like to thank D. G. Blair, L. Ju and C. Zhao
for their keen support of his visit to AEI and Caltech.

\appendix

\section{Necessity of a sub-Heisenberg accuracy for revealing non-classicality}\label{App0}
As we have mentioned in the introduction part, a sub-Heisenberg accuracy is a necessary
condition to probe the non-classicality if the Wigner function of the prepared quantum state has
some negative regions, which do not have any classical counterpart.

To prove the necessity, we use the relation between $Q$ function and Wigner function as pointed out
by Khalili \cite{Khalili1}.
Given density matrix $\hat \rho$, the $Q$ function in the coherent state basis $|\alpha)$
is equal to \cite{Gar2004,Scully,Walls}
\be
Q=\frac{1}{\pi}(\alpha|\hat \rho|\alpha),
\ee
which is always positive defined. It is the Fourier transform of the following characteristic function:
\be
J(\beta, \beta^*)={\rm Tr}[e^{i\beta^*\hat a}e^{i\beta\hat a^{\dag}}\hat \rho].
\ee
Here $\hat a$ is the annihilation operator and is related to the normalized
oscillator position $\hat x/\delta x_q$ and momentum $\hat p/\delta p_q$ [cf. Eq. \eqref{dxq}] by the standard relation
\be
\hat a=[(\hat x/\delta x_q)+i(\hat p/\delta p_q)]/2.
\ee
If we introduce the real and imaginary parts of $\beta$, namely, $\beta=\beta_{\rm r}+i\beta_{\rm i}$, characteristic function $J$
can be rewritten as
\be\label{A4}
J(\beta_{\rm r}, \beta_{\rm i})=e^{-(\beta_{\rm r}^2+\beta_{\rm i}^2)/2}{\rm Tr}[e^{i\beta_{\rm r}(\hat x/\delta x_q)+i\beta_{\rm i} {(\hat p/\delta p_q)}}\,\hat \rho],
\ee
where we have used the fact that $e^{\hat A}e^{\hat B}=e^{\hat A+\hat B}e^{[\hat A,\,\hat B]/2}$, as
$[\hat A,\,\hat B]$ commutes with $\hat A$ and $\hat B$.
Inside the bracket of Eq. \eqref{A4}, it is the characteristic function for the Wigner function $W(x,p)$, and thus
\begin{align}\nonumber
J(\beta_{\rm r}, \beta_{\rm i})&=\frac{1}{(2\pi)^2}\int {dx'dp'}e^{-(\beta_{\rm r}^2+\beta_{\rm i}^2)/2}
\\&e^{-i\beta_{\rm r}(x'/\delta x_q)-i\beta_{\rm i} {(p'/\delta p_q)}}W(x',p').
\end{align}
Integrating over $\beta_{\rm i}$ and $\beta_{\rm i}$, the resulting $Q$ function is given by
\be\label{Q}
Q(x, p)=\frac{1}{2\pi}\int dx'dp' e^{-\frac{1}{2}\left[\frac{(x-x')^2}{\delta x_q^2}+\frac{(p-p')^2}{\delta p_q^2}\right]}W(x', p').
\ee
This will be the same as Eq. \eqref{Wre}, if we identify $W_{\rm recon}(x, p)$ with $Q(x, p)$ and
\be
{\bf V^{\rm add}}=\left[\begin{array}{cc}\delta x_q^2&0\\0&\delta p_q^2\end{array}\right],
\ee
which is a Heisenberg-limited error.
Since squeezing and a rotation of $\hat x$ and $\hat p$ axes will not change the positivity of the $Q$ function,
Eq. \eqref{Q} basically dictates that the reconstructed Wigner function will always be positive if a Heisenberg-limited error is introduced during the verification stage. Therefore, only if a
sub-Heisenberg accuracy is achieved will we be able to reveal the non-classicality of the prepared
quantum state.

\section{Wiener-Hopf method for solving integral equations}\label{AppA}
In this appendix, we will introduce the mathematical method
invented by N.~Wiener and E.~Hopf for solving special type of integral equations. For more details, one
can refer to a comprehensive presentation of this method and its applications by B.~Noble \cite{Noble}. Here we will focus on integral equations that can be brought into the following form as encountered in obtaining the optimal verification scheme:
\begin{equation}
\label{inteq_gen}
\int_0^{+\infty} dt' C(t,t') g(t') =h(t)\,,\quad t>0\,.
\end{equation}
with
\begin{multline}\label{kernelWH}
C(t,t') = A(t-t') +\\ \sum_{\alpha}\int_0^{\min[t,t']} dt'' B^*_{\alpha}(t-t'')B_{\alpha}(t'-t'')\,,
\end{multline}
where $\alpha=1,2,\ldots$ and $B_{\alpha}(t)=0$ if $t<0$.

Assuming that solution to $g(t)$ be a square-integrable function in $\mathcal{L}^2(-\infty,\infty)$, one can split it into \textit{causal} and \textit{anticausal} parts as:
\be
g(t) = g_+(t)+g_-(t)\,,
\ee
where $g_-(t)$ is \textit{causal} part
\be
g_-(t) = \begin{cases}
          0,\ t>0\\
	g(t),\ t\leqslant0
         \end{cases}
\ee
and $g_+(t)$ is the \textit{anticausal} part of $g(t)$
\be
g_+(t) = \begin{cases}
          g(t),\ t>0\\
	0,\ t\leqslant0\,.
         \end{cases}
\ee
This definition enables us to expand the limits of integration in (\ref{inteq_gen}) and (\ref{kernelWH}) to $-\infty<(t,t',t'')<\infty$:
\begin{equation}
\label{inteq_gen2}
\int_{-\infty}^{+\infty} dt'\,C(t,t') g_+(t') =h(t)\,,\quad t>0\,,
\end{equation}
where
\begin{multline}\label{kernelWH2}
C(t,t') = A(t-t') +\\ \sum_{\alpha}\int_{-\infty}^{+\infty} dt'' [B^*_{\alpha,+}(t-t'')B_{\alpha,+}(t'-t'')]_{(+,t'')}\,,
\end{multline}
index $(+,t'')$ stands for taking causal part of a multidimensional function in the argument $t''$.

Let us first exercise the method in a simple special case when $B_\alpha(t)\equiv0,\ \forall\alpha$, this gives a conventional Wiener-Hopf integral equation
\begin{equation}\label{WHeqsimp}
\int_{0}^{+\infty} dt'\,A(t-t') g(t') =h(t)\,,\quad t>0\,,
\end{equation}
which can be rewritten as
\begin{equation}
\left[\int_{-\infty}^{+\infty} dt'\,A(t-t') g_+(t')-h(t)\right]_{(+,t)} =0\,.
\end{equation}
Applying Fourier transform in $t$ and the convolution theorem, one gets:
\be\label{FDWHeq}
\int_{-\infty}^{+\infty}\frac{d\Omega}{2\pi}\Bigl[\tilde A(\Omega)\tilde g_+(\Omega)-\tilde h(\Omega)\Bigr]_+e^{-i\Omega t} = 0\,.
\ee
The spectrum of causal (anticausal) function is simply
\be
\tilde g_{+(-)}(\Omega) = \int_{-\infty}^{\infty} dt\,g_{+(-)}(t)e^{i\Omega t}\,.
\ee
However, this evident relation is not operational for us, as it provides no intuition on how to directly get $\tilde g_{\pm}(\Omega)$ given $\tilde g(\Omega)$ in disposal. The surprisingly simple answer gives complex analysis. Without loss of generality, we can assume that $g(t)$ asymptotically goes to zero at infinity as: $\forall t: |g(t)|<e^{-\gamma_0|t|}$ where $\gamma_0$ is some arbitrary positive number, that guarantees regularity of $\tilde g(\Omega)$ at $-\infty<\Omega<\infty$. In terms of analytic continuation $\tilde g(s)$ of $\tilde g(\Omega)$ to the complex plane $s = \Omega+i\gamma$, the above assumption means that all the poles of $\tilde g(s)$ are located outside its band of analyticity $\,-\gamma_0<\mathrm{Im}(s)<\gamma_0$. Thus, the partition into causal and anticausal parts for $\tilde g(s)$ is now evident:
\be\label{pmpartition}
\tilde g(s) = \tilde g_+(s) +\tilde g_-(s)
\ee
where $\tilde g_+(s)(\tilde g_-(s))$ stands for function equal to $\tilde g(s)$ for $\gamma>\gamma_0(<-\gamma_0)$ and analytic in the half plane above (below) the line $\gamma=\gamma_0(-\gamma_0)$ \footnote{Functions $\tilde g_+(s)$ and $\tilde g_-(s)$ are, in essence, Laplace transforms of $g(t)$ for positive and negative time respectively with only substitution of variable $s\rightarrow ip$.}. According to properties of analytic continuation, this decomposition is unique and completely determined by values of $\tilde g(\Omega)$ on the real axis. Moreover, as a Fourier transform of valid ${\cal L}^2$-function, it has to approach zero when $|s|\rightarrow\infty$. For more general cases, this requirement could be relaxed to demand that $\infty$ should be a regular point of $\tilde g(s)$ so that $\lim\limits_{|s|\rightarrow\infty}\tilde g(s) = const$. This allows to include $\delta$-function and other integrable distributions into consideration, though makes us to add the constant $g(\infty)$ to formula (\ref{pmpartition}) as additional term. For example, for $g(t)=e^{-\alpha|t|}$, $\alpha>0$ one has the following fourier transform:
\be
\tilde g(s) = \frac{2\alpha}{\alpha^2+s^2} = \frac{2\alpha}{(s+i\alpha)(s-i\alpha)}
\ee
that has one pole $s_+=-i\alpha$ in the lower half complex plane (LHP) and one $s_-=+i\alpha$ in the upper half complex plane (UHP). To split $\tilde f(\Omega)$ in accordance with (\ref{pmpartition}) one can use well known formula:
\be\label{resformula}
\tilde g_{\pm}(s) = \sum_{\{s_{\pm,k}\}}\frac{\mathrm{Res}[\tilde g(s),\,s_{\pm,k}]}{(s-s_{\pm,k})^{\sigma_k}}
\ee
where summation goes over all poles $\{s_{+,k}\}$ (with $\sigma_k$ is the order of pole $s_{+,k}$) of $\tilde g(s)$ that belong to the LHP for $\tilde g_+(s)$ and over all poles $\{s_{-,k}\}$ of $\tilde g(s)$ that belong to the UHP for $\tilde g_-(s)$ otherwise, and $\mathrm{Res}[\tilde g(s),s]$ stands for residue of $\tilde g(s)$ at pole $s$. For our example function this formula gives:
\be
\tilde g_+(s) = \frac{i}{s+i\alpha}\,,\quad \tilde g_-(s) = -\frac{i}{s-i\alpha}\,.
\ee
Using the residue theorem, one can easily show that:
\ba
g_+(t) &=& e^{-\alpha t},\mbox{ for }t>0\\
g_-(t) &=& e^{\alpha t},\mbox{ for }t<0\,.
\ea

Coming back to the equation (\ref{FDWHeq}), assume that function $\tilde A(\Omega)$ can be factorized in the following way:
\be
\tilde A(\Omega)=\tilde a_-(\Omega)\tilde a_+(\Omega)
\ee
where $\tilde a_{+(-)}(\Omega)$ is a function analytic in the UHP (LHP) with its inverse, \i.e., both its poles and zeroes are located in the LHP (UHP). One gets the following equation:
\be\label{eqp}
\left[\tilde a_-(\Omega)\tilde a_+(\Omega)\tilde g_+(\Omega)-\tilde h(\Omega)\right]_+=0\,.
\ee
To solve this equation, one realizes the following fact:
for any function $\tilde f$, $[\tilde f(\Omega)]_+=0$ means that $\tilde f$ has no poles in the LHP. Multiplication of $\tilde f$ by any function $\tilde g_-$ which also has no poles in the LHP will evidently not change the equality, namely, $[\tilde g_-(\Omega)\tilde f(\Omega)]_+=0$. Multiplying Eq. \eqref{eqp} by $1/\tilde a_-(\Omega)$, the solution reads
\be
\tilde g_+(\Omega) = \frac{1}{\tilde a_+(\Omega)}\Bigl[\frac{\tilde h(\Omega)}{\tilde a_-(\Omega)}\Bigr]_+\,.
\ee
Performing inverse Fourier transform of $\tilde g_+(\Omega)$, the time-domain solution $g_+(t)$ can be obtained.

Now we are ready to solve Eq. \eqref{inteq_gen2} with the general kernel in Eq. \eqref{kernelWH2}. Performing similar manipulations, one obtain the following equation for $\tilde g_+(\Omega)$ in the Fourier domain:
\begin{equation}
\left[\left(\tilde A +\sum_\alpha\tilde B_{\alpha} \tilde B_{\alpha}^*\right)\tilde g_+ - \sum_\alpha \tilde B_{\alpha}(\tilde B^*_{\alpha} \tilde g_+)_- - \tilde h\right]_+=0\,,
\end{equation}
where we have omitted arguments $\Omega$ of all functions for brevity. Since $\tilde B_{\alpha}$ is a causal function, $\tilde B^*_{\alpha}$ is anticausal and $\tilde g_+$ is causal,  $( \tilde B^*_{\alpha} \tilde g)_-$ only depends on the value of $\tilde g$ on the poles of $\tilde B^*_{\alpha}$. Performing similar factorization
\begin{equation}
\tilde \psi_+ \tilde \psi_- = \tilde A + \sum_{\alpha} \tilde B_{\alpha}\tilde B_{\alpha}^*
\end{equation}
with $\tilde\psi_+$ ($\tilde\psi_-$) and $1/\tilde\psi_+$ ($1/\tilde\psi_-$) analytic in the UHP (LHP), $\psi_+(-\Omega) = \psi_+^*(\Omega) = \psi_-(\Omega)$, we get the solution in the form:
\begin{equation}
\label{gtilde}
\tilde g_+ = \frac{1}{\tilde \psi_+}\left[\frac{\tilde h}{\tilde \psi_-}\right]_+
+ \frac{1}{\tilde \psi_+}\left[\sum_{\alpha} \frac{\tilde B_{\alpha} (\tilde B^*_{\alpha} \tilde g_+)_-}{\tilde \psi_-}\right]_+\,.
\end{equation}
Even though $\tilde g_+$ also enters the right hand side of the above equation, yet $(\tilde B^*_\alpha \tilde g_+)_-$ can be written out explicitly as:
\be
(\tilde B^*_{\alpha} \tilde g_+)_- = \sum_{\{\Omega_{-,k}\}}\frac{\tilde g_+(\Omega_{-,k})\mathrm{Res}[\tilde B^*(\Omega),\,\Omega_{-,k}]}{(\Omega-\Omega_{-,k})^{\sigma_k}}.
\ee
Here $\{\Omega_{-,k}\}$ are poles of $\tilde B^*(\Omega)$ that belong to UHP, and therefore $\tilde g_+(\Omega_{-,k})$ are just constants that can be obtained by solving a set of linear algebra equations evaluating Eq. \eqref{gtilde} at those poles $\{\Omega_{-,k}\}$.

\section{Solving integral equations in Section \ref{III}}\label{AppB}
Here we will use the technique introduced in the previous section to obtain analytical solutions to the integral equations we encountered in the subsections \ref{IIIC} and \ref{IIID}.

In the coordinate representation, the integral equations for $g_{1,2}$ are the following [cf. Eqs. \eqref{52} and \eqref{53}]:
\begin{align}
\int_0^{T_{\rm int}}dt'\left[\begin{array}{cc}{\bf C}_{11}(t, t')& {\bf C}_{12}(t,t')\\
{\bf C}_{21}(t,t')&{\bf C}_{22}(t,t')\end{array}\right]
\left[\begin{array}{c}g_1(t')\\g_2(t')\end{array}\right]
=\left[\begin{array}{c}0\\h(t)\end{array}\right],
\end{align}
where ${\bf C}_{ij}\,(i,j=1,2)$ are given by Eq. \eqref{32} and we have defined $h(t)\equiv \mu_1f_1(t)+\mu_2 f_2(t)$. Since the optimal
$g_{1,2}(t)$ will automatically cut off when $t>\tau_F$, we can extend
the integration upper bound $T_{\rm
int}$ to $\infty$. It brings those equations into the right shape considered in the Appendix \ref{AppA}. In the frequency domain, they can be written as
\begin{align}
&[\tilde S_{11}\tilde g_1]_++[\tilde{S}_{12}\,\tilde g_2]_+=0,\\
&[\tilde{S}_{21}\,\tilde g_1]_++[\tilde S_{22}\,\tilde g_2]_+-\tilde\Gamma=\tilde h,\\
&\tilde\Gamma=(1-\eta)(\Omega_q^4/2)(e^{2q}+2{\zeta_F^2})[\tilde G_x (\tilde G_x
\tilde g_2)_-]_+.
\end{align}
Here $\tilde S_{ij}$ are the Fourier transformation
of the correlation functions ${\bf C}_{ij}$. Specifically, they are
\begin{align}
\tilde S_{11}&=\frac{\eta+(1-\eta)e^{2q}}{2},\\
\tilde S_{12}&=-\frac{(1-\eta)e^{2q}\Omega_q^2}{2(\Omega+\omega_m-i\gamma_m)(\Omega-\omega_m-i\gamma_m)},\\
\tilde S_{21}&=\tilde S^{*}_{12},\\
\tilde S_{22}&=\frac{\Lambda^2}{4}+
\frac{(1-\eta)(e^{2q}+2{\zeta_F^2})\Omega_q^4}{2[(\Omega+\omega_m)^2+\gamma_m^2][(\Omega-\omega_m)^2+\gamma_m^2]}.
\end{align}

Since $\tilde S_{11}$ is only a number, the solution to $\tilde g_1$ is simply
\be
\tilde g_1=-\tilde S_{11}^{-1}[\tilde S_{12}\tilde
g_2]_+.
\ee
In the time-domain, this recovers the result in Eq. \eqref{59}.
Through a spectral factorization
\be
\tilde \psi_+\tilde\psi_-\equiv\tilde S_{22}-\tilde
S_{11}^{-1}\tilde S_{12}\tilde S_{21},
\ee
we obtain the solution for $\tilde g_2$:
\begin{align}
\tilde g_2=&\frac{1}{\tilde \psi_+}\left\{\frac{1}{\tilde
\psi_-}\left[\tilde h-\tilde
S^{-1}_{11}\tilde S_{21}(\tilde
S_{12}\tilde g_2)_-+\tilde\Gamma\right]\right\}_+.
\end{align}
Plugging $\tilde \Gamma$ into the above equation, $\tilde g_2$ becomes
\be \tilde
g_2=\frac{1}{\tilde \psi_+}\left\{\frac{1}{\tilde \psi_-}\left[\tilde h+\kappa\,\tilde
G_x(\tilde G_x^*\tilde
g_2)_-\right]\right\}_+
\ee with $\kappa\equiv
m^2\Omega_q^4{\zeta_F'^2}$.  A simple inverse Fourier transformation gives
$g_1(t)$ and $g_2(t)$. The unknown Lagrange multipliers can be solved using
Eq. \eqref{50}. We can then derive the covariance matrix ${\bf V}^{\rm add}$
for the added verification noise with Eq. \eqref{Vadd}. In the free-mass regime,
a closed form for ${\bf V}^{\rm add}$ can be obtained as shown explicitly in Eq. \eqref{covmat}.

\end{document}